\documentclass[final]{article}

\usepackage{arxiv}

\usepackage[utf8]{inputenc} 
\usepackage[T1]{fontenc}    
\usepackage{hyperref}       
\usepackage{url}            
\usepackage{booktabs}       
\usepackage{amsfonts}       
\usepackage{nicefrac}       
\usepackage{microtype}      
\usepackage{lipsum}
\usepackage{graphicx}

\usepackage{xcolor}
\usepackage{tikz}
\usepackage{amsmath}
\usepackage{amssymb}

\usepackage{lineno}
\usepackage{algorithm2e}
\usepackage{textcomp} 
\usepackage{subcaption}
\usepackage{adjustbox}

\modulolinenumbers[5]
\usetikzlibrary{calc}
\usetikzlibrary{intersections}
\usetikzlibrary{through}

\usepackage{nomencl}
\usepackage{etoolbox}

\makenomenclature

\title{Comparative Analysis of Methods for Cloud Segmentation in Ground-Based Infrared Images}

\author{
 Guillermo Terr\'en-Serrano \\
  Department of Electrical and Computer Engineering \\
  The University of New Mexico \\
  Albuquerque, NM 87131, United States\\
  \texttt{guillermoterren@unm.edu} \\
 \And
  Manel Mart\'inez-Ram\'on \\
  Department of Electrical and Computer Engineering \\
  The University of New Mexico \\
  Albuquerque, NM 87131, United States\\
  \texttt{manel@unm.edu} \\
}

\begin{document}

\maketitle

\begin{abstract}
    The increasing penetration of photovoltaic systems in the power grid makes it vulnerable to cloud shadow projection. Real-time cloud segmentation in ground-based infrared images is important to reduce the noise in intra-hour global solar irradiance forecasting. We present a comparison between discriminative and generative models for cloud segmentation. The performances of supervised and unsupervised learning methods in cloud segmentation are evaluated. The discriminative models are solved in the primal formulation to make them feasible in real-time applications. The performances are compared using the j-statistic. Infrared image preprocessing to remove stationary artifacts increases the overall performance in the analyzed methods. The inclusion of features from neighboring pixels in the feature vectors leads to a performance improvement in some of the cases. Markov Random Fields achieve the best performance in both unsupervised and supervised generative models. Discriminative models solved in the primal yield a dramatically lower computing time along with high performance in the segmentation. Generative and discriminative models are comparable  when preprocessing is applied to the infrared images.
\end{abstract}


\printnomenclature[2cm]

\nomenclature[A]{\textbf{PV}}{Photovoltaic}
\nomenclature[A]{\textbf{IR}}{Infrared}
\nomenclature[A]{\textbf{GSI}}{Global solar irradiance}
\nomenclature[A]{\textbf{DAQ}}{Data acquisition system}
\nomenclature[A]{\textbf{FOV}}{Field of view}
\nomenclature[A]{\textbf{CNN}}{Convolutional neural networks}
\nomenclature[A]{\textbf{MRF}}{Markov random field}
\nomenclature[A]{\textbf{ICM}}{Iterated conditional modes}
\nomenclature[A]{\textbf{SVC}}{Support vector classifier}
\nomenclature[A]{\textbf{RRC}}{Ridge regression for classification}
\nomenclature[A]{\textbf{NBC}}{Naive Bayes classifier}
\nomenclature[A]{\textbf{GDA}}{Gaussian discriminant analysis}
\nomenclature[A]{\textbf{GMM}}{Gaussian mixture model}
\nomenclature[A]{\textbf{SA}}{Simulated annealing}
\nomenclature[A]{\textbf{MALR}}{Moist adiabatic lapse rate}
\nomenclature[A]{\textbf{WLK}}{Weighted Lucas-Kanade}
\nomenclature[A]{\textbf{ML}}{Maximum likelihood}
\nomenclature[A]{\textbf{MAP}}{Maximum a posteriori}
\nomenclature[A]{\textbf{EM}}{Expectation-maximization}
\nomenclature[A]{\textbf{ROC}}{Receiver operating characteristic}
\nomenclature[A]{\textbf{LOO}}{Leave-one-out}
\nomenclature[A]{\textbf{VSH}}{Voting scheme}
\nomenclature[A]{\textbf{GPC}}{Gaussian process for classification}

\section{Introduction}

A large power grid system fully operated using only renewable power in Europe can be theoretically possible by 2050 \cite{ZAPPA2019}, and the projected solar energy share will increase to 47\% by 2050 in the USA \cite{EIA2021}. Clouds increase or decrease the global solar irradiance reaching Earth's surface. This is of great importance when a considerable percentage of the energy in a power grid is generated using large Photovoltaic (PV) systems \cite{CHEN2020}. Even when the PV arrays in a power plant are arranged in a configuration capable of attenuating the effects caused by moving clouds, cloud shadows produce interruptions in energy generation which may be out of the grid operator's admissible range \cite{LAPPALAINEN2017}. Moving clouds have an effect not only on the generation of energy from PV systems, but also on solar thermal power plants \cite{CRESPI2018}. The inclusion of cloudiness information from sky images into a statistical model for forecasting Global Solar Irradiance (GSI) improves the overall performance of the prediction \cite{FURLAN2012}.

Computer recognition of clouds is a geospatial information problem \cite{Smith2007}. The tropopause limits the range of cloud formations, which seasonally varies across latitudes \cite{Randel2000}. Different cloud types are expected to be found at a different range of altitudes \cite{Houze2014}. When using features extracted from color intensity channels, cloud patterns inferred from data acquiesced at different latitudes may not be correlated. Feature extraction methods based on Gabor filter texture analysis and statistics are more easily replicable across databases \cite{Deng2019}.

In GSI forecasting \cite{Hong2020}, ground-based methods without features extracted from clouds are not effective in intra-hour forecasting and are generally used when the forecasting horizon is hours-ahead \cite{Perez2010, Bouzerdoum2013, Mellit2014, Salcedo2014, Lauret2015, GARCIA2018}. For an intra-day hours-ahead forecast, the effects of clouds on ground-level solar irradiance can be assessed using satellite images \cite{JIANG2020, PRASAD2015} and the accuracy of numerical weather models for GSI nowcasting is improved when cloudiness information is extracted from satellite images \cite{Mathiesen2013}. Ground measurements (GSI or PV power) are an option for intra-day forecasting, but are not capable of predicting ramp events (when a cloud will abruptly affect a PV system \cite{Ichi2017}). Ground-based sky imaging is the most suitable method in applications requiring intra-hour GSI forecasting \cite{DIAGNE2013, SOBRI2018, PAZIKADIN2020}. 

When using visible light ground-based sensors, the circumsolar region appears saturated in images including the Sun \cite{Fu2013, Shi2019}. Structures that block the Sun's direct irradiance partially obstruct the images \cite{Chow2011, Dev2017, Li2015, Ye2019}, creating forecasting problems \cite{Yang2018}. Nevertheless, total sky imagery \cite{Gohari2014, Marquez2013}, and fisheye lenses are capable of recording a large Field of View (FOV) \cite{KONG2020}. When these technologies are applied with the aim of motion estimation, fisheye lens' distortion should be removed \cite{Cheng2017a}. Recent ground-based Infrared (IR) sky imaging developments increase the FOV of thermal images \cite{Redman2018, MAMMOLI2019}. IR images allow for the derivation of physical features of the clouds such as temperature \cite{Escrig2013} and height, which are more interpretable for modelling physical processes. Ground-based IR cameras providing radiometric measures \cite{SHAW2005, SHAW2013}, are used to study statistical cloud features \cite{THURAIRAJAH2007}, and in earth-space communications applications \cite{NUGENT2009}. Further research shows how to stabilize the thermal image from microbolometers in atmospheric measurements \cite{NUGENT2013}. 

Previous investigations in cloud segmentation concluded that pixel segmentation using features extracted from neighboring pixels improves performance \cite{Hu2015, Shi2017}. Graph models based on neighboring pixels' classification are referred to as MRFs. They are a generalization of the Ising Model, first introduced in  ferromagnetic problems \cite{ISING1925}, and later applied to 2-dimensional crystal lattice problems \cite{LARS1944}. The Iterated Conditional Modes (ICM) algorithm, developed for unsupervised training of MRFs in image processing \cite{BESAG1986}, was implemented for IR satellite image cloud segmentation  \cite{Papin2002}, and visible light ground-based images \cite{Li2012}.

The superpixel approach speeds-up computing time, but produces a coarse segmentation \cite{Liu2015}. Real-time cloud segmentation is a problem for kernel learning methods, as the Gram matrix is generally dense \cite{Taravat2015, Zhuo2014}. One alternative is the use of primal formulation optimization. The same problem appears with Convolutional Neural Networks (CNN). The required computing time is high \cite{Deng2019}, although it is considerably reduced when using GPUs \cite{Dronner2018, Zhang2018}. Nevertheless, these methods require data augmentation and regularization techniques to avoid overfitting. Otherwise, the conclusions obtained are not comparable between different databases of cloud images, since the distribution of the features will vary. We prove that when effective preprocessing is applied to the IR images to extract informative physical features, discriminative models are faster and have similar accuracy to generative, kernel or CNN methods.

This research utilizes data acquired from an innovative radiometric long-wave IR sky-imaging system, rather than a visible light sky-imaging system. The Data Acquisition (DAQ) system is mounted on a solar tracker. An advantage of using the IR sky-imaging system is that the saturated circumsolar region of the image is smaller. The saturation of the circumsolar area removes necessary information about the clouds for intra-hour GSI forecasting. Another advantage of the IR imaging system is that thermal images allow for the extraction of physical features that are useful for cloud segmentation. A novelty of this work is the implementation of a preprocessing algorithm to increase the cloud segmentation performances in IR sky images. The proposed preprocessing algorithm applies two models to the IR images \cite{TERREN2021}. The first model reproduces the scattering effect caused by debris (e.g. water stains and dust) on the outdoor germanium window of the camera. The second model reproduces the effect of direct irradiation from the Sun and scatter irradiation from the atmosphere to remove saturation in the circumsolar region, making it possible to differentiate between the Sun and clouds. 

Cloud segmentation is useful to identify which pixels in an image are cloudy and which are clear-sky. This information can then be used in a solar forecasting algorithm. This research contributes to the field of cloud segmentation and solar forecasting through a comparative analysis of generative and discriminative models. The objective is to determine which model performs better in an IR sky-imaging system mounted on a solar tracker. The discriminative methods used in the analysis are: Ridge Regression for Classification (RRC) \cite{SHAWE2004}, Support Vector Classifier (SVC) \cite{burges1998tutorial}  and Gaussian Processes \cite{Rasmussen2006} for Classification (GPC). The training and testing time is drastically reduced when the RRC, SVC and GPC models are implemented in their primal formulation, because the number of dimensions obtained after mapping data to the Hilbert space is much smaller compared to the dual formulation. MRFs are part of the analyzed generative models. MRF models are computationally expensive but suitable for segmentation problems, because information from the classification of neighboring pixels is included in the prior. The generative models include effective methods with low computational requirements. The training and testing computation time is improved by simplifying the covariance matrix. The Naive Bayes Classifier (NBC) and k-means clustering are simplifications of the Gaussian Discriminant Analysis (GDA) and Gaussian Mixture Model (GMM) respectively. The performances of generative models are compared between supervised (NBC, GDA and MRF) and unsupervised learning algorithms (k-means, GMM and ICM-MRF). Unsupervised learning models are less time intensive because they do not require labels to train a segmentation model, simplifying training for the user. The Simulated Anneling (SA) algorithm is implemented to perform an intelligent optimization that reduces the testing time of the MRF and ICM-MRF. A voting scheme improves the overall cloud segmentation performance of an algorithm \cite{Hu2015}. In this investigation, the performances of the voting schemes that use all proposed methods and the optimal combination of methods are compared.  

\section{Datasets and Measurements}

The features from the IR images are extracted after processing to remove the Sun and atmospheric scattering effects. Due to difficulty accessing the DAQ location, the IR camera's window cannot be routinely cleaned. We implemented a persistent model to remove the effects produced by dust particles and water spots on the window, using standard weather parameters from a nearby weather station \cite{TERREN2021}, prior to the feature extraction. The extracted features from a pixel and its neighboring pixels are cross-validated to find the set of features that increases the segmentation performances in each method.

\subsection{Data Acquisition System}

The proposed segmentation methods utilize a DAQ system equipped with a solar tracker that updates its pan and tilt every second, maintaining the Sun in a central position in the images throughout the day. The IR sensor is a Lepton\footnote{https://www.flir.com/} radiometric camera with an 8 to 14 $\mu m$ wavelength. The pixels in a frame are emission temperature measurements in centi-kelvin units. Henceforth, the emission temperature is referred to as ``temperature'' for short. The resolution of an IR image is $80 \times 60$ pixels, and the diagonal FOV is $60^\circ$. The DAQ is located on the roof of the UNM-ME building in Albuquerque, NM. The data is publicly accessible in the Dryad repository \cite{GIRASOL}.

The weather parameters that were used to compute cloud height and remove cyclostationary artifacts on the IR images are: atmospheric pressure, air temperature, dew point and humidity. The weather station performs a measurement every 10 minutes. The data was interpolated to match the IR images samples. The weather station is located at the University of New Mexico Hospital. It is publicly accessible\footnote{https://www.wunderground.com/dashboard/pws/KNMALBUQ473}.

\subsection{Feature Extraction}

A pixel of the camera frame is defined by a pair of Euclidean coordinates $i,j$. The temperature of the clouds in an IR image are $\mathbf{T} = \{ T_{i,j} \in \mathbb{R}^+ \mid \forall i = 1, \ldots, M, \ \forall j = 1, \ldots, N \}$ in Kelvin, and are measured using the radiometric IR camera. The temperature of a particle in the troposphere is roughly a function of the height \cite{Hess1959}, so that the height of a pixel in a frame can be approximated using the Moist Adiabatic Lapse Rate (MALR) function \cite{Stone1979}, that we define as $\phi : (T^{air}, T^{dew}, P^{atm}) \mapsto \Gamma_{MARL}$ \cite{TERREN2021}. The weather parameters necessary to compute $\Gamma_{MARL}$ are: air temperature $T^{air}$, dew point $T^{dew}$ and atmospheric pressure $P^{atm}$ (measured on ground-level). The height of a cloudy pixel is computed using this formula $H_{i,j} = [T_{i,j} - T^{air}] / \Gamma_{MARL}$. The height of the pixels are defined in kilometers as $\mathbf{H} = \{ H_{i,j} \in \mathbb{R}^+ \mid \forall i = 1, \ldots, M, \ \forall j = 1, \ldots, N \}$ and they are computed using the MALR function.

\subsection{Image Preprocessing}

The DAQ system germanium outdoor lens is cleaned when it rains. However, after the water droplets have evaporated, a water stain is left on the lens. Due to the inconvenience of cleaning the lens in person, we propose implementing a model of the stains caused by dried water droplets. The algorithm to model the outdoor lens begins with a classification model determining the sky conditions in an IR image. 
    
A linear SVC model is trained to distinguish between four classes of sky-conditions: clear-sky, cumulus, stratus or nimbus cloud. The feature vectors of the model include the mean, variance, kurtosis and skewness of the temperatures $\mathbf{T}_{i,j}$ in the images, and the clear-sly index values. When sky conditions are detected as clear-sky, the IR image is added to the clear-sky set. At the same time, the algorithm forgets the oldest clear-sky image in the set. The clear-sky set contains the last $L = 250$ of clear-sky images. The scatter irradiance produced by dust and water stains on the outdoor germanium lens is the median image computed using all the IR images in the clear-sky set $\mathbf{W} = \{ w_{i,j} \in \mathbb{R} \mid \ \forall i = 1, \ldots, M, \ \forall j = 1, \ldots, N \}$. The outdoor lens algorithm and the sky conditions classification model are fully detailed in \cite{TERREN2021}. In this investigation, we use the obtained temperature of each pixel, after removing the dust and stains. The temperatures are defined as $\mathbf{T}^\prime = \{ T^\prime_{i,j} \in \mathbb{R}^+ \mid \ \forall i = 1, \ldots, M, \ \forall j = 1, \ldots, N \}$, and the heights are $\mathbf{H}^{\prime} = \{ H_{i,j}^\prime \in \mathbb{R}^+ \mid \forall i = 1, \ldots, M, \ \forall j = 1, \ldots, N \}$.

The image preprocessing method implemented in this article is introduced in \cite{TERREN2021}. The raw temperature of a pixel $i,j$ is processed using a model that combines the effects of the scatter irradiance $\mathcal{S}(\cdot)$ and the direct irradiance from the Sun $\mathcal{D}(\cdot)$. The function of the background atmospheric irradiance $\mathcal{A}(\cdot)$ is,
\begin{equation}
    \label{eq:atmosperic_irradiance}
    \begin{split}
        \mathcal{A} \left(i,j; \mathbf{x}_0, \boldsymbol{\Theta} \right) &= \mathcal{S} \left( j; y_0, \theta_1, \theta_2 \right) + \mathcal{D} \left( i,j; \mathbf{x}_0, \theta_3, \theta_4 \right) \\
        &= \theta_1 \exp \left\{\frac{j - y_0}{\theta_2}\right\} + \theta_3 \left\{ \frac{ \theta_4^2}{\left[ \left( i - x_0 \right)^2 + \left( j - y_0 \right)^2 + \theta_4^2\right]^{\frac{3}{2}}} \right\}
        \end{split}
\end{equation}
where the parameters are $\boldsymbol{\Theta} = \{ \theta_1, \dots , \theta_4 \}$ and $\mathbf{x}_0 = \{x_0, y_0\}$ is the position of the Sun in the image. This function is used to model the deterministic component of the irradiance in the IR images. 

The optimal parameters of the scatter irradiance model are different in each image. We propose to model these parameters as the predictors of a function whose covariates are: air temperature, dew point, and the Sun's elevation and azimuth angle. To approximate the parameters of the modeling function, the atmospheric background model is optimized in a set of frames with clear-sky conditions from different days of year. The parameters of the Sun's direct radiation model are constants. 

After removing both the window model and the atmospheric model from the images, differences of temperature with respect to the tropopause's temperature are defined as $\mathbf{\Delta T} = \{ \Delta T_{i,j} \in \mathbb{R} \mid \forall i = 1, \ldots, M, \ \forall j = 1, \ldots, N \}$. The differences of height are also computed and multiplied by the tropopause's average temperature in the image, estimated using the atmospheric background model. The resulting heights are $\mathbf{H}^{\prime \prime} = \{ H_{i,j}^{\prime \prime} \in \mathbb{R}^+ \mid \forall i = 1, \ldots, M, \ \forall j = 1, \ldots, N \}$. 
    
The temperature differences are normalized to 8 bits, $\mathbf{I} = \{ i_{i,j} \in \mathbb{N}^{2^{8}} \mid \forall i = 1, \ldots, M, \ \forall j = 1, \ldots, N \}$. The aim of the normalization is to extract a feature that simplifies the classification of a model. Through the normalization, information about the feasible minimum temperature of a cloud is added to each pixel. The normalization formula is $i_{i,j} = [ \Delta T_{i,j} - \mathbf{min} ( \Delta \mathbf{T}) ]/[(11.5 - 1.52) \cdot 9.8]$, the lowest value is set to 0, and then divided by the clouds' maximum feasible temperature \cite{TERREN2021}. The feasible temperature is calculated assuming a linear temperature decrease of $9.8 \mathrm{K/km}$ in the tropopause \cite{Hummel1981}, and that the average tropopause height is $11.5 \mathrm{km}$ at $36^\circ$ latitude north \cite{Pan2011}. The average height above sea level is $1.52\mathrm{km}$ in Albuquerque, NM.

The velocity vectors were computed applying the Weighted Lucas-Kanade method (WLK) \cite{TERREN2020b, SIMON2003}. For each two consecutive images ${\bf I}^{k-1}$, ${\bf I}^{k}$ of the data set, the velocity vectors are defined as $\mathbf{V}^k = \{ \mathbf{v}_{i,j} = (u,v)^k_{i,j} \in \mathbb{R}^2 \mid \ \forall i = 1, \ldots, M, \ \forall j = 1, \ldots, N \}$. The upper index $k$ denoting the frame is omitted in the rest of the document.

The pixels in the images that form the dataset were manually labelled as clear $y_{i,j} = 0$ or cloudy $y_{i,j} = 1$. The temperature in the background of the images varies. The background temperature is the temperature of the tropopause. For each image, this temperature was first identified, and then used to distinguish pixels that have the image background temperature. 

\subsection{Feature Vectors}

To find the optimal feature combination, we propose the validation of different physical features extracted from a pixel, and three sets of neighboring pixels, included as dependent variables in the model. 

The first feature vector, $\mathbf{x}^{1}_{i,j} = \{ T_{i,j}, \ H_{i,j} \}$, contains the raw radiometric temperature of the pixels and the heights computed using the raw temperatures. The second feature vector, $\mathbf{x}^{2}_{i,j} = \{ T_{i,j}^\prime, \ H_{i,j}^\prime \}$, contains the temperature and height of the pixels after removing the artifacts on the IR camera's window. The third feature vector, $\mathbf{x}^{3}_{i,j} = \{ \Delta T_{i,j}, \ H_{i,j}^{\prime \prime} \}$, contains the incremental temperatures and heights after removing the Sun's direct radiation and the atmospheric scatter radiation. The fourth feature vector includes the magnitude of the velocity vectors, the normalized increments of temperature, and the increments of temperature; and is defined as $\mathbf{x}^{4}_{i,j} = \{ \mathrm{mag} (\mathbf{v}_{i,j}), \ i_{i,j}, \ \Delta T_{i,j} \}$.

To segment a pixel, its feature vectors and those of its neighboring pixels are introduced into the classifier. In the experiments, we define \emph{$1^{st}$ order neighborhood} feature vector as the set of four pixels closest to the test pixel $i,j$, \emph{$2^{nd}$ order neighborhood} is defined as the eight closest pixels, and term \emph{single pixel} is used when no neighbors are included, that is: 
\begin{itemize}
    \item Single pixel: 
            $\{ \mathbf{x}_{i,j} \}, \quad \forall i,j = i_1,j_1, \ldots, i_{M}, j_{N}$
    \item $1^{st}$ order neighborhood: $\{\mathbf{x}_{i-1,j}, \ \mathbf{x}_{i,j - 1}, \ \mathbf{x}_{i,j + 1}, \ \mathbf{x}_{i+1,j} \}$.
            
    \item $2^{nd}$ order neighborhood: $\{\mathbf{x}_{i-1,j}, \ \mathbf{x}_{i,j - 1}, \ \mathbf{x}_{i,j + 1}, \ \mathbf{x}_{i+1,j}, \ \mathbf{x}_{i - 1,j -1}, \ \mathbf{x}_{i - 1,j + 1}, \ \mathbf{x}_{i + 1,j + 1}, \ \mathbf{x}_{i + 1,j + 1} \}$.
\end{itemize}

\section{Methods}

The methods summarized below can be classified as generative when they have the capacity of generating new samples from a likelihood model, that is, when the model implements a density approximation of the form $p ( {\bf x} | \mathcal{C}_k )$ where $\mathcal{C}_k $ is the segmentation label of the pixel. Discriminative models do not have the ability to generate data since they implement a direct approximation of the posterior $p ( \mathcal{C}_k  | {\bf{x}} )$.

\subsection{Generative Models}

Generative models are either Maximum Likelihood (ML), or Maximum A Posteriori (MAP) methods. When generative models use an input feature structure, together with the use of an energy function for the probabilistic modeling of data (Ising model), they are generally known as MRF models. We summarize below the discriminant analysis, which applies ML inference, GMM and k-means clustering, and supervised and unsupervised MRF methods, with MAP inference.

\subsubsection{Discriminant Analysis}

GDA and NBC are both supervised learning methods, because the training dataset input features $\mathbf{x}_i$ are paired with a label $\mathcal{C}_k$. As we assume that the prior in these models is uniform, the inference applied is ML.

\paragraph{Gaussian Discriminant Analysis} 

GDA obtains the posterior probability of $y_i = \mathcal{C}_k$ given a set of features $\mathbf{x}_i \in \mathbb{R}^d$ when applying the Bayes theorem \cite{HASTIE2001}, where a prior is chosen over the classes, and a Gaussian likelihood is used for the observations. The posterior of class $\mathcal{C}_k$, where $k \in \{1, \ldots, K \}$ are possible classes, is maximized by the Bayes' rule with the expression  ${ p \left( \mathbf{x}_i \right) } \propto p \left( \mathcal{C}_k \right) p \left( \mathbf{x}_i \mid \mathcal{C}_k \right)$.

The corresponding means $\boldsymbol{\mu}_k \in \mathbb{R}^d$ and covariance matrices  $\boldsymbol{\Sigma}_k \in \mathbb{R}^{d \times d}$ are estimated with the samples that have assigned class $\mathcal{C}_k$ and $d$ is the sample dimension, i.e, the number of features in vector $\mathbf{x}_i$.

\paragraph{Naive Bayes Classifier}

The NBC applies the Bayes theorem, similarly to a ML classifier, but it computes a likelihood by assuming that all features are independent. Therefore, the corresponding Gaussian likelihood is approximated by a product of a univariate Gaussian distribution per each dimension of the observation $\bf x$ \cite{MURPHY2012}.  

\subsubsection{Clustering}

The GMM and k-means are unsupervised learning algorithms. Their respective objective functions group the samples in clusters represented by conditional likelihood functions, and then a posterior distribution for each class $\mathcal{C}_k$ is computed with the likelihood and a prior distribution of the labels. Thereby, the inference level applied is MAP. K-means can be considered as a simplification of the GMM.

\paragraph{Gaussian Mixture Model}

The GMM assumes a known number $K$ of possible latent variable values. For each one, a Gaussian distribution is constructed. Initial values are proposed for these parameters and for the priors of the classes. The Expectation Maximization (EM) algorithm \cite{MURPHY2012} is used to iteratively adjust all these parameters. In the E step, posteriors $p(\mathcal{C}_k|{\bf x}_i)$ are computed for all samples. In the M step,  mean $\mu_k$ is computed by averaging all samples weighted by their corresponding posterior. The covariance is computed similarly. The priors are computed by averaging the posteriors. Once these statistics are computed, the E step is repeated, until convergence.

\paragraph{k-means}

K-means is a simplification of a GMM, where covariances  $\mathbf{\Sigma}_k = \mathbf{I}_{d \times d}$, are assumed to be constant \cite{FORGY1965, MURPHY2012}. The posterior distribution for the latent variables is 1 for the class whose mean is closer to the sample and zero otherwise.

\subsubsection{Markov Random Fields}

In a MRF, pixels are grouped in cliques of a graph $\mathcal{G}$ where a clique is a a set of nodes which are neighbours of each other given a definition of neighborhood. For the problem at hand, a node represents a pixel, and a neighborhood is defined in terms of pixel proximity. The prior probability of a pixel's class is a normalized exponential of an energy function. This represents the \emph{energy} of a clique. 

The energy function of a MRF is composed of  $\varphi$, or the joint distribution of a class, and  $\psi$, the potential energy of the system's configuration \cite{STAN2001}
\begin{align}\label{eq:energy_function1}
    \mathcal{E} \left( y_i, \mathbf{x}_i \right) = \sum_{i} \varphi\left(\mathbf{x}_i, y_i \right) + \sum_{i,j} \psi \left( y_i, y_j \right),
\end{align}
In graph $G$, a sample $i$ has a set of neighboring pixels, and each neighboring sample $j$ has class $y_j$.  A sample $\mathbf{x}_i$ is classified using the Bayes' theorem,
\begin{equation}\label{eq:Bayes_theorem}
    p \left( y_{i} = \mathcal{C}_k \mid \mathbf{x}_{i}, \boldsymbol{\theta}_k \right) 
    \propto p \left( \mathbf{x}_{i} \mid y_{i} = \mathcal{C}_k, \boldsymbol{\theta}_k  \right) p \left( y_{i} = \mathcal{C}_k \right).
\end{equation}
where   $p \left( \mathbf{x}_{i} \mid y_{i} = \mathcal{C}_k, \boldsymbol{\theta}_k  \right)= \mathcal{N} (\boldsymbol{\mu}_k, \boldsymbol{\Sigma}_k)$ with   $\boldsymbol{\theta}_k = \{\boldsymbol{\mu}_k, \boldsymbol{\Sigma}_k \}$ . The log-likelihood of class $\mathcal{C}_k$ is defined as $\varphi\left(\mathbf{x}_i, y_i \right) $ in the energy function \eqref{eq:energy_function1}. The prior is 
\begin{equation}\label{eq:potential_function}
    p \left( y_{i} \right) = \frac{1}{Z} \exp \left( - \psi \left( y_{i}  \right) \right)  = \frac{1}{Z} \exp \left( \sum_{i,j \in \Omega_\ell} y_i \beta y_j \right),
\end{equation}
where the potential function has been factorized in cliques of a graph $\mathcal{G}$ by applying the Hammersley–Clifford theorem \cite{HAMMERSLEY1971}. A clique is defined as a set of nodes that are all neighbors of each other \cite{MURPHY2012}. With the above model, a posterior can be constructed to classify the pixels. If there are no labelled images, then unsupervised inference of the class parameters  in a MRF model can be performed using the ICM algorithm \cite{BESAG1986}. The standard optimization goes through all the pixels calculating their potential and classifying them in each iteration of the algorithm. The computational cost of this method is high, but we can assume that it is not necessary to evaluate the pixels whose state has high energy, because their classification will not change. The computational cost can be reduced by sampling the pixels that are likely to be misclassified, and applying the optimization procedure only to them.

We propose to optimize the configuration of the pixels in an IR image applying the SA algorithm \cite{KIRK1983} to the MRF models \cite{KATO2001}. The SA algorithm is applied on the implementation, after the inference of the class distributions. Temperature parameter $T$ in the SA algorithm is linearly cooled down $T^{t + 1} = \alpha T^{t}$ at each iteration $t$ with an acceptance rate $\alpha$. The optimal parameter $\alpha$ is a trade off between accuracy and speed.

\subsection{Discriminative Models}

The discriminative algorithms in this work are based on kernel models \cite{SHAWE2004}, where the data is implicitly transformed into a Hilbert space $\mathcal{H}$ of higher dimensionality (possibly infinite) with a dot product expressable as a positive definite function of the input data  \cite{aizerman1964}. Kernel Learning uses the generalized Representer Theorem \cite{scholkopf2001}, which states that, under mild conditions, any machine can be represented by a linear combination of dot products between the training and test data (which is called a \emph{dual} formulation). This implies the use of a matrix containing the $N^2$ kernel dot products between data, which can be overwhelming due to the high quantity of data to be used during the training in the present problems. Therefore, we use explicit basis functions for the nonlinear transformation into $\mathcal{H}$. The matrices to be manipulated have the same dimension of this space.  

The transformation is a polynomial expansion of the elements of  $\bf x$.  The  expansion is defined as $\varphi : \mathcal{X} \mapsto \mathcal{P}^n$, where $n$ is the order of the expansion and $\mathcal{P}^n =  [ ( n + d) ! ] / n!$. 
The polynomial expansion of the dataset $\mathcal{D} = \{ \mathbf{\Phi}, \ \mathbf{y} \}$, is defined in matrix form as
\begin{equation}
    \mathbf{\Phi} = \left[ 
    \begin{array}{ccc}
    \varphi \left( \mathbf{x}_1 \right) &
    \cdots &
    \varphi \left( \mathbf{x}_N \right)
    \end{array} \right] \in \mathbb{R}^{\mathcal{P}^n \times N}, \quad \mathbf{y} = 
    \begin{bmatrix}
    y_1 \\
    \vdots \\
    y_N
    \end{bmatrix}, 
\end{equation}
where $y_i \in \{ 0, \ 1\}$ which are labels for a clear or cloudy pixel, respectively. The polynomial expansion  is used in the primal formulated kernel for RR, SVC and GPC is defined such as
\begin{align}
    \begin{split}
        \varphi \left( {\bf x}_i \right) &= \left[ 1 ~\cdots~ a_j x_j ~\cdots~ a_{j,k} x_j x_k ~\cdots~ a_{j,k,l} x_j x_k x_l ~\cdots~ \right]^\top \in \mathbb{R}^{\mathcal{P}^n}, \\
        & \quad \quad \quad \forall j,k,l \cdots = 1, \dots, D
    \end{split}
\end{align}
where scalars $a_j, ~a_{j,k},~a_{j,k,l}, ~\cdots \in \mathbb{R}$ are chosen so that the corresponding dot product in the space can be written as
\begin{align}
	\varphi\left( \mathbf{x}_i \right)^\top \varphi \left(\mathbf{x}_i\right ) = \left[ 1 + \mathbf{x}_i^\top \mathbf{x}_i \right]^n.
\end{align} 
which is the well known polynomial kernel of order $n$. 

\subsubsection{Ridge Regression}

The RRC is a minimum mean squared error method with quadratic norm regularization applied on the parameters $\mathbf{w}$,
\begin{align}
    \label{eq:ridge_regression}
    \min_{\mathbf{w}} \ \sum_{i = 1}^{N} \left( \mathbf{y} - \mathbf{w}^\top \mathbf{\Phi} \right)^2 + \gamma \| \mathbf{w} \|_2.
\end{align}
where $\gamma$ is the regularization parameter, and it requires cross validation. The solution is simply 
$ \bar{\mathbf{w}} = ( \mathbf{\Phi}\mathbf{\Phi}^\top + \gamma  \mathbf{I} )^{-1} \mathbf{\Phi}^\top \mathbf{y}.$
The output is passed through a sigmoid function in order to provide it with probability mass properties, i.e   
\begin{equation}
    p \left( y = 1 \mid \varphi \left( \mathbf{x} \right), \mathcal{D} \right) = \frac{1}{1 + \exp \left( - \mathbf{\bar{w}}^\top \varphi \left( \mathbf{x} \right) \right)}
\end{equation}

\subsubsection{Primal solution for Support Vector Machines}

Several approaches to solve the SVC in the primal space have been proposed, as the iterative re-weighted least squares method \cite{Navia2001}, or by directly solving the following quadratic problem \cite{FAN2008, HSU2010}, 
\begin{align}
    \label{eq:linear_svm}
    \min _{\mathbf{w}} \frac{1}{2} \| \mathbf{w}\|_2+C \sum_{i=1}^{N} \left( \max \left[ 0,1-y_i \mathbf{w}^\top \varphi \left( \mathbf{x}_i \right)  \right] \right)^2,
\end{align}
where $C > 0$ is a penalty term, $x_i({\bf w},{\bf x}_i,y_i)$ is the loss function, and in our model the norm is a $L_2$. This is a maximum margin problem \cite{FAN2008}. The complexity parameter $C$ has to be cross-validated. 

The original formulation of the linear SVC does have not a probabilistic output. However, the distance from a sample to the hyper-plane can be transformed to a probability measure,
\begin{equation}
    \begin{split}
    p \left( \mathcal{C}_1 \mid \varphi \left( \mathbf{x}_* \right), \mathcal{D}  \right) &= \frac{1}{1 + \exp \left( - \mathbf{\bar{w}}^\top \varphi \left( \mathbf{x}_* \right) \right)} \\
    p \left( \mathcal{C}_2 \mid \varphi \left( \mathbf{x}_* \right), \mathcal{D}  \right) &= 1 - p \left( \mathcal{C}_1 \mid \varphi \left( \mathbf{x}_* \right), \mathcal{D} \right),
    \end{split}
\end{equation}
using the sigmoid function, similarly to the proposed RRC for classification.

\subsubsection{Primal solution for  Gaussian Processes}

When formulated in the primal, a GPC is is known as Bayesian logistic regression \cite{MURPHY2012, Rasmussen2006, Jaakkola1997}. A posterior over the primal parameters $\bf w$ is computed as
\begin{align}
    p \left( \mathbf{w} \mid \mathcal{D} \right) \propto  p \left( \mathbf{y} \mid \mathbf{\Phi}, \mathbf{w} \right)  p \left( \mathbf{w} \right).
\end{align}
The likelihood function of the model is $ p ( y_i \mid \mathbf{\Phi}, \mathbf{w} ) = \prod_{i = 1}^N \hat{y}_i^{y_i} ( 1 - \hat{y} )^{1 - y_i}$, where $\mathbf{\hat{y}} = [ \hat{y}_1 \dots \hat{y}_N ]^\top$ are the predictions. The prior is Gaussian $ p ( \mathbf{w} ) \sim \mathcal{N} \left( \mathbf{w}  \mid \boldsymbol{\mu}_0, \mathbf{\Sigma_0} \right)$, but the posterior is not Gaussian.  In contrast to Bayesian linear regression, this approach does not have an analytical solution.  The Laplace approximation is applied to assume that the posterior is Gaussian $ q ( \mathbf{\bar{w}} ) \sim \mathcal{N} ( \mathbf{w} \mid \mathbf{\bar{w}}, \mathbf{\Sigma_n} )$. The optimal set of parameters are found maximizing the marginal log-likelihood via numerical gradient.

\section{J-Statistic}

The Younde's j-statistic or Younde's Index is a test to evaluate the performances of a binary classification \cite{YOUDEN1950}, that is defined as,
\begin{align}
    J = sensitivity + specificity - 1.
\end{align}
The entries on the confusion matrix are used to compute the sensitivity,
\begin{align}
    sensitivity = \frac{TP}{TP + FN},
\end{align}
where $TP$ and $FN$ are the true positives and false negatives, and the specificity is,
\begin{align}
    specificity = \frac{TN}{TN + FP},
\end{align}
where $TN$ and $FP$ are the true negatives and false positives. It is different than the accuracy score of a binary classification, which is also obtained using the entries of the confusion matrix, and that it is,
\begin{align}
    ACC = \frac{TP + TN}{TP + FP + TN + FN}.
\end{align}

As the optimized loss function is different in each model, we propose to define a prior $\lambda$, which has to be cross-validated for each one of models, and has an optimal value for each classification function,
\begin{equation}
    \begin{split}
    \label{eq:virtual_prior}
    p \left( \mathcal{D} \mid \mathcal{C}_k \right) &= \frac{p \left( \mathcal{C}_k \mid \mathcal{D} \right) p \left( \mathcal{C}_k  \right)}{p \left( \mathcal{D} \right)} \\ 
    &\propto p \left( \mathcal{C}_k \mid \mathcal{D}  \right) p \left( \mathcal{C}_k \right) \\
    &\propto p \left( \mathcal{C}_k \mid \mathcal{D}  \right) \lambda
    \end{split}
\end{equation}
so the maximized loss function is the same in all the models. The classification probabilities are defined as $p \left( \mathcal{D} \mid \mathcal{C}_1 \right) = p \left( \mathcal{C}_1 \mid \mathcal{D}  \right) \lambda $, and $ p \left( \mathcal{D} \mid \mathcal{C}_2 \right) = 1 - p \left( \mathcal{D} \mid \mathcal{C}_1 \right)$. The j-statistic score is maximized finding the optimal binary classification $\lambda$ threshold. For that, the j-statistic is applied to the conventional Receiver Operating Characteristic (ROC) analysis \cite{FAWCETT2006}, and it is computed at each point of the ROC. We propose to use the maximum value of j-statistic in the ROC curve as the optimal point.

After the cross-validation of the virtual prior $\lambda$, a class $\mathcal{C}_k$ is assigned to a sample $\mathbf{x}_*$ following this criteria,
\begin{align}
    \hat{y}_* = \underset{k}{\operatorname{argmax}} \ p \left( \mathcal{C}_k \mid \mathbf{x}_*, \mathcal{D} \right) \lambda,
\end{align}
which is a MAP estimation.

\section{Experiments}

The selected samples constitute the dataset used in the cross-validation and testing of the segmentation models. The samples include different days in all four seasons. The sample images include partially cloudy, fully clear-sky and fully covered sky conditions. The images were captured at different hours of the day, so the Sun's elevation and azimuth angle are different. Therefore, the atmospheric background model is different in all of the images. The dataset is composed of different types of clouds found at different heights in the tropopause. We found that the most difficult clouds for the models to classify are cirrus stratus, which are also included in the dataset. Artificially created clouds like contrails are also included in testing set. Contrails are highly difficult for the models, as the scattering effect is similar to that produced by cirrus clouds. The dataset is composed of 12 images with labels, amounting to a total of 57,600 pixels. They are organized chronologically and divided into training (earlier dates) and testing set (later dates). The training set has 7 images, which are 33,600 pixels in total. The testing set has the remaining 5 images, which are 24,000 pixels. The training set contains 5 images with clouds, 1 image with clear-sky, and another one with covered sky conditions. The testing set has 3 images with clouds, 1 with clear-sky, and 1 with covered sky conditions.

\begin{figure}[!ht]
    \centering
    \includegraphics[scale = 0.325]{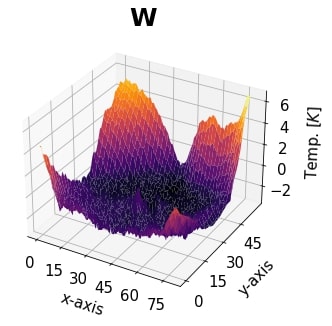}
    \includegraphics[scale = 0.325]{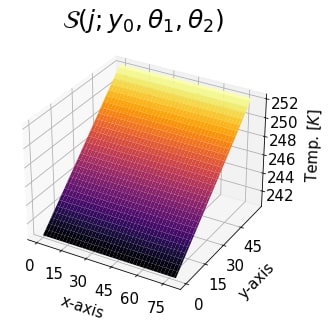}
    \includegraphics[scale = 0.325]{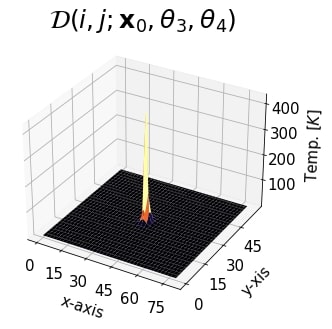}
    \includegraphics[scale = 0.325]{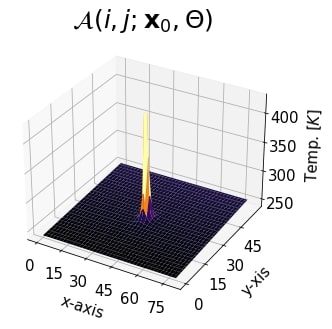}
\caption{Models applied to preprocess the radiometric IR images. The models correspond to the IR image in Fig. \ref{fig:features_extraction_steps}. The first image is the atmospheric scattered irradiance. The second is the Sun's direct irradiance. The third is the atmospheric background model that combines both previous models. The last image is the outdoor lens scattering effect model.}
\label{fig:solar_irradiance_models}
\end{figure}

\begin{figure}[!ht]
    \begin{subfigure}{\linewidth}
        \centering
        \includegraphics[scale = 0.45]{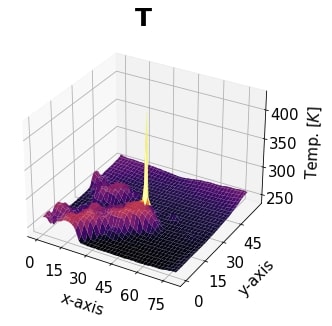}
        \includegraphics[scale = 0.45]{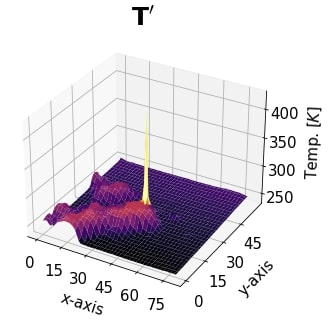}
        \includegraphics[scale = 0.45]{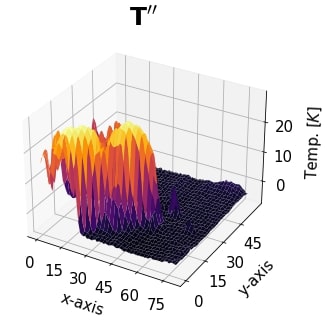}
    \end{subfigure}
    \begin{subfigure}{\linewidth}
        \centering
        \includegraphics[scale = 0.325]{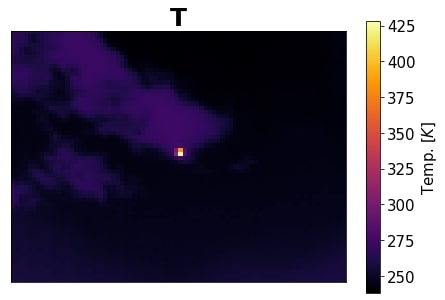}
        \includegraphics[scale = 0.325]{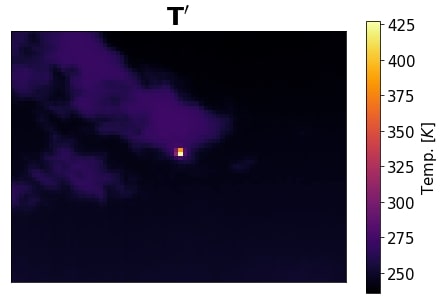}
        \includegraphics[scale = 0.325]{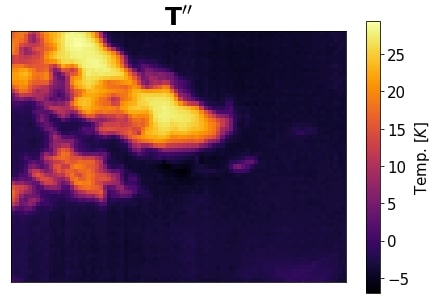}
    \end{subfigure}
\caption{The IR images are shown in a 3-dimensional surface graph in the top row and in the figures in the bottom row. The figures in the first column are the raw pixels obtained from the radiometric IR camera. The figures in the second column show the processed IR images, removing the atmospheric background model. The figures in the third column are the IR images after processing to remove the atmospheric background and the outdoor lens scattering effect. The applied models are shown in Fig. \ref{fig:solar_irradiance_models}.}
\label{fig:features_extraction_steps}
\end{figure}

\begin{figure}[!ht]
    \centering
    \begin{subfigure}{0.325\linewidth}
        \includegraphics[scale = 0.325]{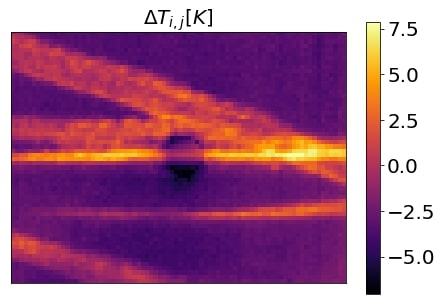}
    \end{subfigure}
    \begin{subfigure}{0.325\linewidth}
        \includegraphics[scale = 0.325]{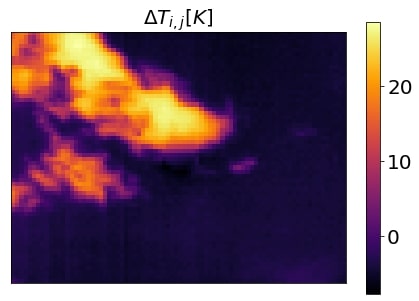}
    \end{subfigure}
    \begin{subfigure}{0.325\linewidth}
        \includegraphics[scale = 0.325]{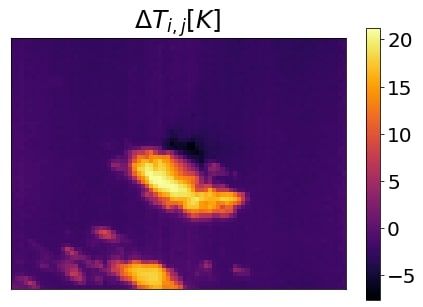}
    \end{subfigure}
    \begin{subfigure}{0.325\linewidth}
        \includegraphics[scale = 0.325]{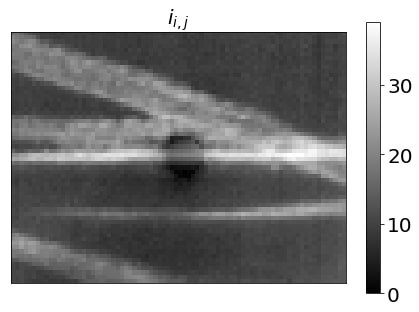}
    \end{subfigure}
    \begin{subfigure}{0.325\linewidth}
        \includegraphics[scale = 0.325]{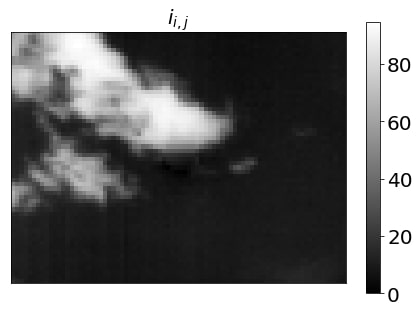}
    \end{subfigure}
    \begin{subfigure}{0.325\linewidth}
        \includegraphics[scale = 0.325]{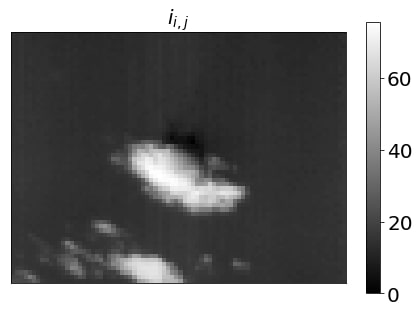}
    \end{subfigure}
    \begin{subfigure}{0.325\linewidth}
        \includegraphics[scale = 0.325]{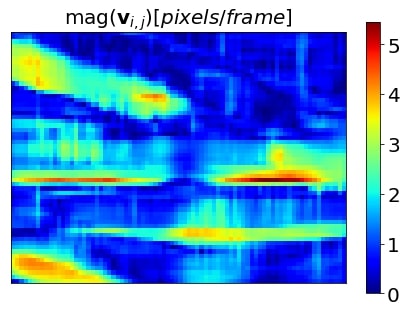}
    \end{subfigure}
    \begin{subfigure}{0.325\linewidth}
        \includegraphics[scale = 0.325]{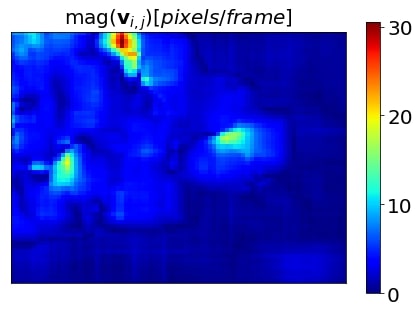}
    \end{subfigure}
    \begin{subfigure}{0.325\linewidth}
        \includegraphics[scale = 0.325]{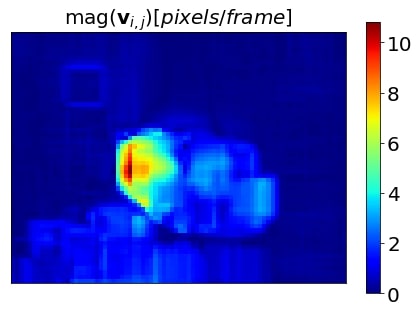}
    \end{subfigure}
    \begin{subfigure}{0.325\linewidth}
        \includegraphics[scale = 0.325]{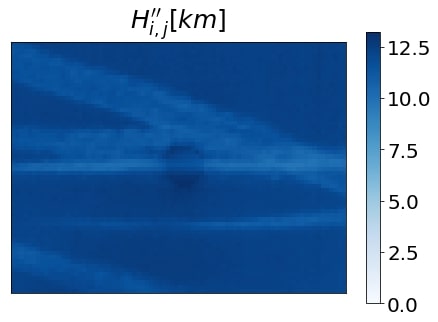}
    \end{subfigure}
    \begin{subfigure}{0.325\linewidth}
        \includegraphics[scale = 0.325]{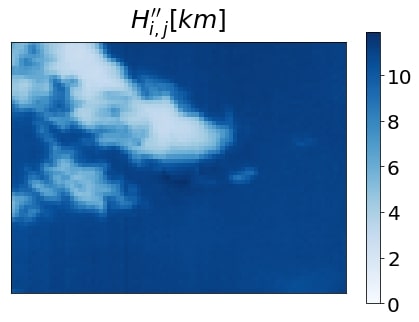}
    \end{subfigure}
    \begin{subfigure}{0.325\linewidth}
        \includegraphics[scale = 0.325]{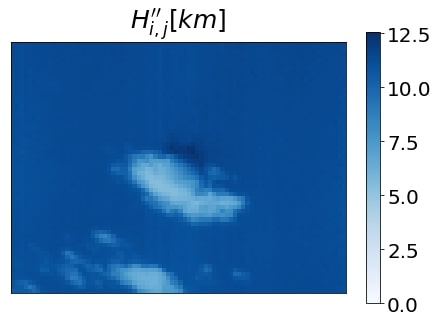}
    \end{subfigure}
\caption{This figure shows the features extracted from three test images. The test images are organized in columns. The images in the first row show the normalized intensity of the pixels. The images in the second row show the magnitude of the velocity vectors. The images in the third row show the increments of temperature with respect to the height of the tropopause. The images in the fourth row show the height of the clouds. The last row shows the test images in which the clouds were manually segmented.}
\label{fig:features}
\end{figure}

\begin{table}[htb!]
\centering
\tiny
\setlength{\tabcolsep}{6.pt} 
\renewcommand{\arraystretch}{1.5} 
\begin{tabular}{lcccc}
\toprule
 & Type of Cloud & Cloud Covered [\%] & Elevation [$^\circ$] & Azimuth [$^\circ$] \\
\midrule
Train No. 1 & Stratocumulus and Cumulus &             38.67 & 29.69 & 160.91 \\
Train No. 2 & Stratocumulus & 28.13 & 28.70 & 157.49 \\
Train No. 3 &  Cirrocumulus and Stratocumulus & 36.6 & 24.43 & 146.99 \\
Train No. 4 & Altocumulus & 4.5 & 31.40 & 183.09 \\
Train No. 5 & Cumulus & 37.94 & 73.52 & 172.20 \\
Train No. 6 & Nimbus & 100 & 29.92 & 164.81 \\
Train No. 7 & Clear-Sky & 0 & 28.94 & 158.59 \\
Test No. 1  & Contrail & 38.67 &  48.76 & 183.94 \\
Test No. 2  & Cumulus & 28.13 & 37.53 & 149.34 \\
Test No. 3  & Altocumulus & 12.29 & 32.10 & 204.17 \\
Test No. 4  & Clear-Sky & 0 &  76.58 & 190.5 \\
Test No. 5  & Altostratus & 100 & 60.07 & 165.62 \\
\bottomrule
\end{tabular}
\caption{Type of clouds, percentage of cloud covered and the Sun's position in the horizon in each IR image of the training and test sets.}
\label{tab:type_of_Clouds}
\end{table}

The Leave-One-Out (LOO) method is implemented in the cross-validation of the parameters. In this method, the training samples are left out for validation one at a time, while the rest of the training samples are used to fit the model. In our problem, the training samples are the images in the training set, so a training image is used for validation while the others are used for training the model. 

The cross-validation is done using a high performance computer. Each validation sample (in the LOO routine) runs on a different CPU, and 7 CPUs are necessary for each experiment. When the LOO routine is finished, the results are communicated to the main node, and a new set of hyperparameters and virtual prior $\lambda$ are validated. This procedure is repeated until all possible combinations of hyperparameters and virtual priors are validated. The LOO routine runs in multiple experiments at the same time. Each experiment has a combination of feature vectors, neighborhoods, polynomial expansions (in the discriminative models) and cliques (in the MRF models). All CPUs are operating at full capacity and are only inactive during the waiting time (i.e. until all jobs of the LOO routine are finished).
    
The cross-validation is computationally expensive due to the amount of training samples, but running the LOO routine and the experiments in parallel reduces the training time by several orders of magnitude. The testing times are obtained when running each segmentation model in a single CPU.
    
Exploratory results showed that the features that work best are those in vectors $\mathbf{x}^{3}_{i,j}$ and $\mathbf{x}^{4}_{i,j}$. All possible combinations were tested, but none produced any improvement in the classification performance, with the exception of those in $\mathbf{x}^{4}_{i,j}$. However, the original features require preprocessing to achieve reasonable performances. This is shown in the classification results obtained by $\mathbf{x}^{1}_{i,j}$ and $\mathbf{x}^{2}_{i,j}$

In the generative models, NBC and k-means clustering do not have hyperparameters. The GDA and GMM have the covariance matrix regularization term $\gamma$ which has to be cross-validated. In the k-means clustering, the feature vectors were standardized $\bar{\mathbf{x}}_{i,j} = [ \mathbf{x}_{i,j} - \mathbf{E} (\mathbf{X}) ] / \mathbf{Var} (\mathbf{X})$. The rest of the models neither required normalization nor standardization of the feature vectors.

\begin{figure}[!htb]
    \begin{minipage}{\linewidth}
    \centering
    \includegraphics[scale = 0.375]{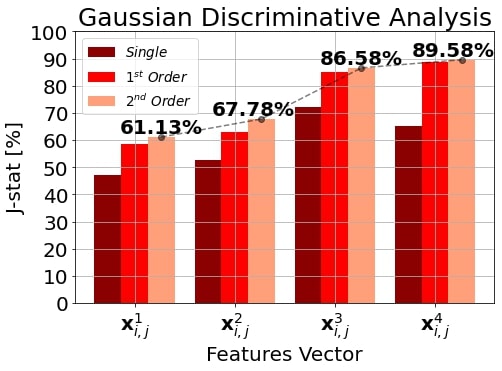}
    \includegraphics[scale = 0.375]{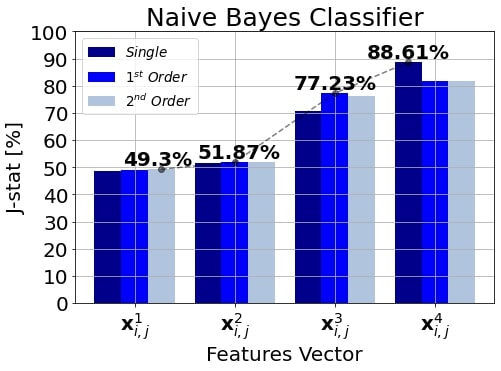}
    \end{minipage}
    \begin{minipage}{\linewidth}
    \centering
    \includegraphics[scale = 0.375]{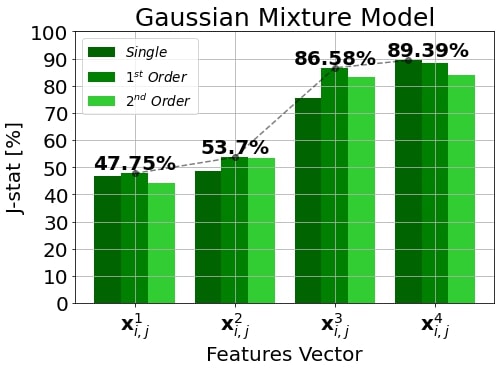}
    \includegraphics[scale = 0.375]{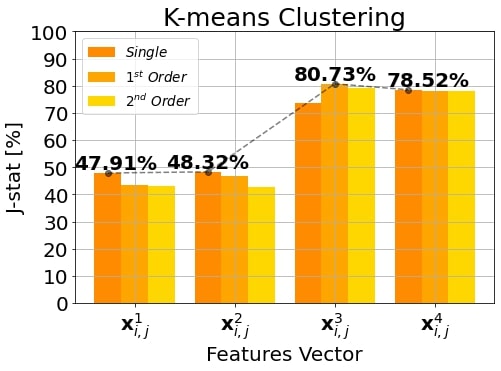}
    \end{minipage}
    \caption{The graph shows the j-statistic achieved by the generative models. The color of the bars in the graph indicate the order of neighborhood from dark to light. The neighborhoods are organized from the left to right within the groups of bars. This corresponds with the order of the feature vectors used in the model.}
    \label{fig:generative}
\end{figure}

In the MRF models, the cliques potential $\beta$ in Eq. \eqref{eq:potential_function} was cross-validated in all the models. The supervised MRF have the covariance matrix regularization term $\gamma$ which was cross-validated. The unsupervised ICM-MRF is computationally expensive, so the regularization term of the covariance matrix was set to $\gamma = 1$. In the supervised MRF with SA in the implementation, the cross-validated parameters were the regularization term of the covariance matrix $\gamma$, and the cooling parameters $\alpha$. In the unsupervised MRF trained with the ICM algorithm (using the SA algorithm in the implementation), the parameters of the regularization term of the covariance matrix and cooling were set to $\gamma = 1$ and $\alpha = 0.75$.

\begin{figure}[!htb]
    \begin{minipage}{\linewidth}
    \centering
    \includegraphics[scale = 0.375]{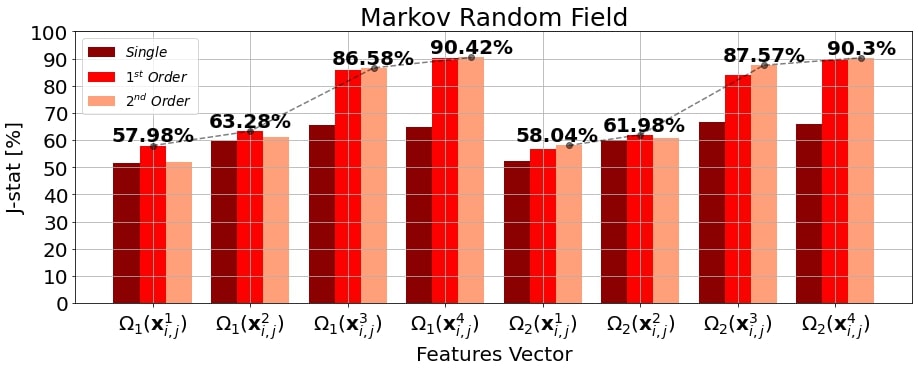}
    \end{minipage}
    \begin{minipage}{\linewidth}
    \centering
    \includegraphics[scale = 0.375]{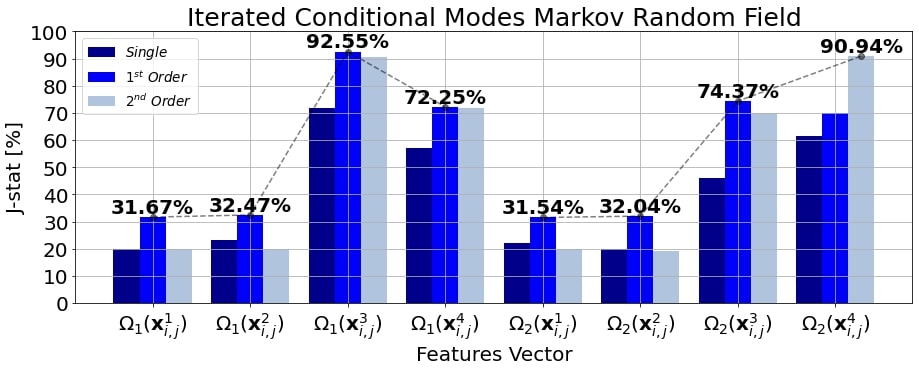}
    \end{minipage}   
    \begin{minipage}{\linewidth}
    \centering
    \includegraphics[scale = 0.375]{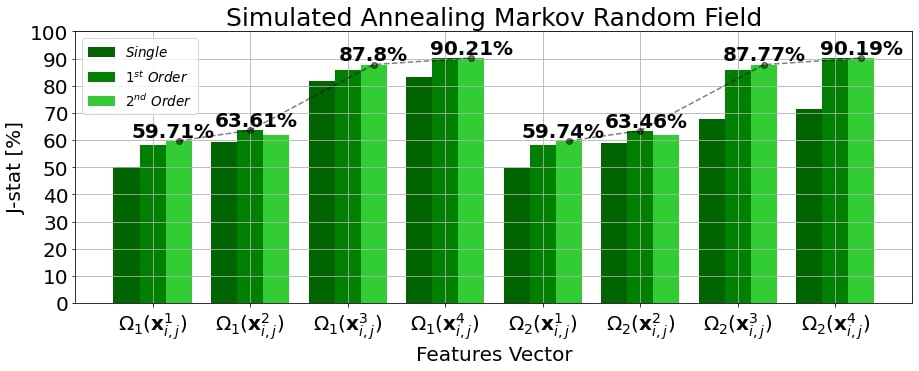}
    \end{minipage} 
    \begin{minipage}{\linewidth}
    \centering
    \includegraphics[scale = 0.375]{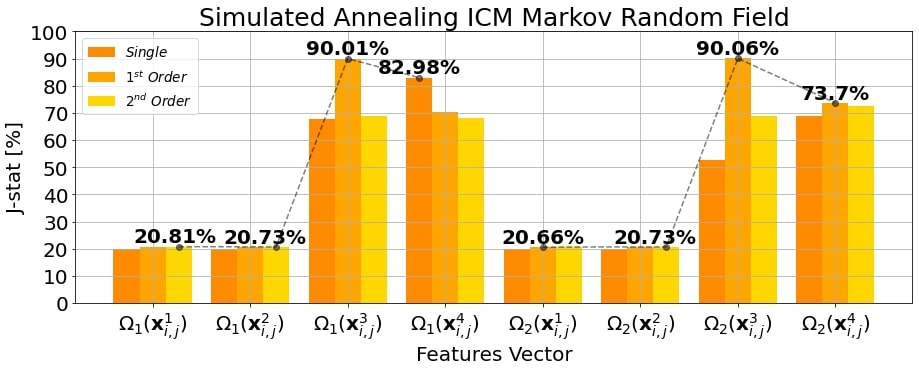}
    \end{minipage} 
\caption{The graphs show the j-statistic achieved by the MRFs using different cliques in their potential function. The four feature vectors are organized in groups of three bars. There are two groups of feature vectors: those with a potential function of $1^{st}$ order cliques $\Omega_1 (\cdot)$, and those with a potential function of $2^{nd}$ order cliques $\Omega_2 (\cdot)$.}
\label{fig:mrf}
\end{figure}

In the discriminative models, the RRC has the regularization $\gamma$ in Eq. \eqref{eq:ridge_regression} that has to be cross-validated. The SVC has the complexity term $C$ of the loss function in Eq. \eqref{eq:linear_svm}. The hyperparameters of the GPC are the prior mean $\boldsymbol{\mu}_0$ and the covariance matrix $\boldsymbol{\Sigma}_0$. The prior mean and covariance matrix are simplified to $\boldsymbol{\mu}_0 \triangleq \mathbf{0}$ and $\boldsymbol{\Sigma}_0 \triangleq \gamma \mathbf{I}_{D \times D}$, so only the parameter $\gamma$ is cross-validated.

\begin{figure}[!htb]
    \begin{minipage}{\linewidth}
    \centering
    \includegraphics[scale = 0.375]{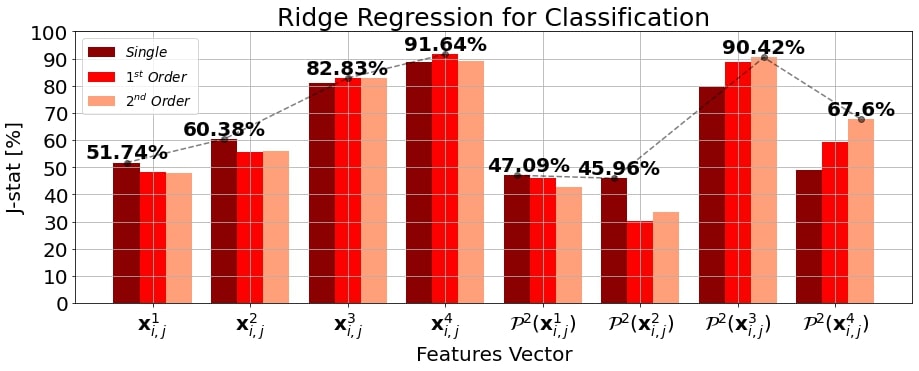}
    \end{minipage}
    \begin{minipage}{\linewidth}
    \centering
    \includegraphics[scale = 0.375]{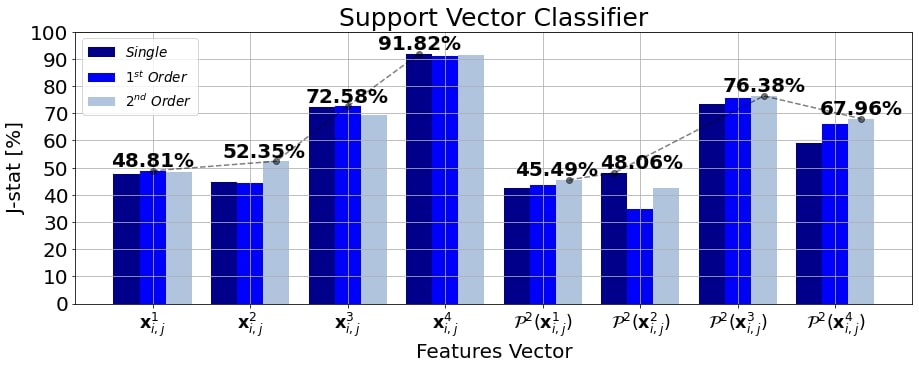}
    \end{minipage}   
    \begin{minipage}{\linewidth}
    \centering
    \includegraphics[scale = 0.375]{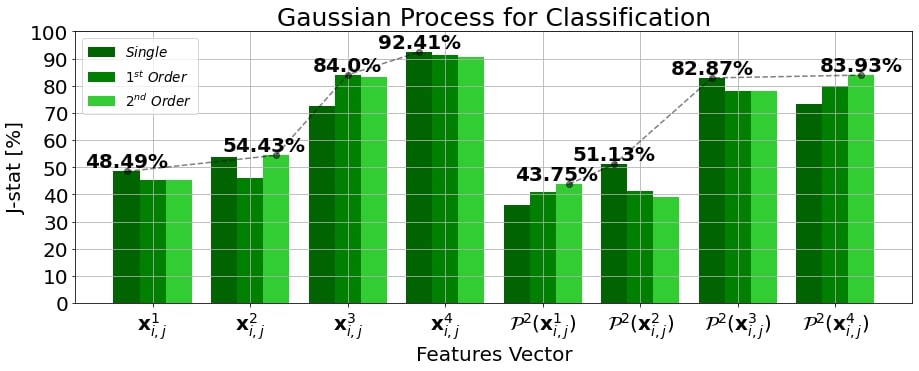}
    \end{minipage} 
\caption{The graphs show the j-statistics achieved by the discriminative models. The feature vectors are organized in groups. The bars in the same group from dark to light are: features extracted from a single pixel, a $1^{st}$ order neighborhood and a $2^{nd}$ order neighborhood. When a polynomial expansion of the second order is applied to the feature vectors, it is denoted as $\mathcal{P}^2 (\cdot)$.}
\label{fig:discriminative}
\end{figure}

In addition to each set of hyperparameters, all models have a virtual prior $\lambda$ that corrects possible class-imbalances in Eq. \eqref{eq:virtual_prior}. The hyperparameters and the virtual prior $\lambda$ have to be cross-validated. A set of hyperparameters define the ROC curve, and the virtual prior $\lambda$ is used to find the optimal j-statistic along this curve with the predicted probabilities of each class for each combination. The validation j-statistic is the average of the j-statistics obtained in each LOO cross-validation loop. The model selection criteria is the highest validation j-statistic. 

The experiments were carried out in the Wheeler high performance computer of the UNM-CARC, which uses a SGI AltixXE Xeon X5550 at 2.67GHz with 6 GB of RAM memory per core, 8 cores per node, 304 nodes total, and runs at 25 theoretical peak FLOPS. Linux CentOS 7 is installed.

\section{Discussion}

\begin{figure}[htb]
    \centering
    \begin{minipage}{\linewidth}
    \includegraphics[scale = 0.375]{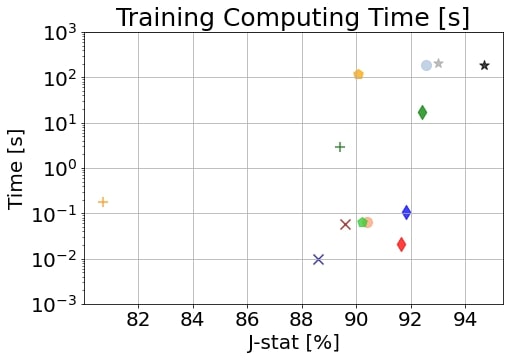}
    \includegraphics[scale = 0.375]{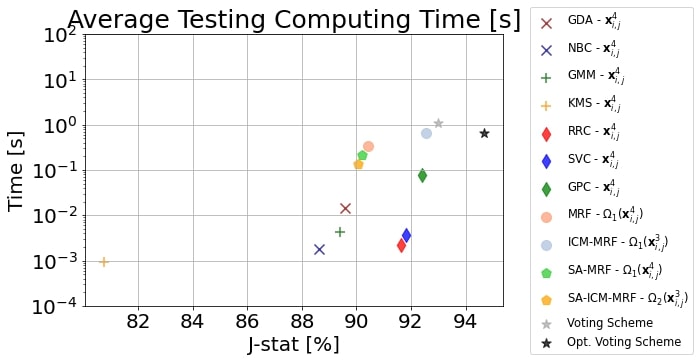}
    \end{minipage}
\caption{Left: Computing time of each model during training. Right: Average computing time during the segmentation in the test subset. The legend displays the optimal feature vectors, neighborhood order, polynomial expansion and cliques of each model.}
\label{fig:timing}
\end{figure}

\begin{figure}[!htb]
    \begin{subfigure}{\linewidth}
        \centering
        \includegraphics[scale = 0.35]{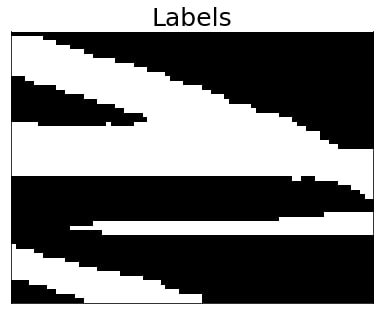}
        \includegraphics[scale = 0.35]{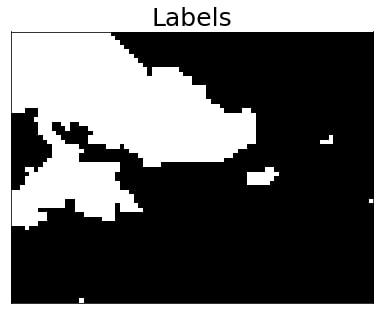}
        \includegraphics[scale = 0.35]{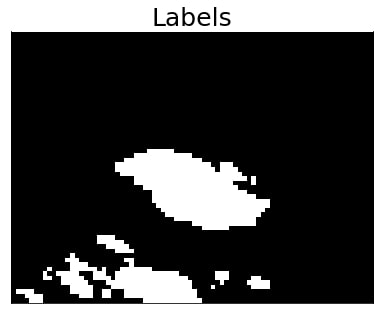}
    \end{subfigure}
    \begin{minipage}{\linewidth}
        \centering
        \includegraphics[scale = 0.35]{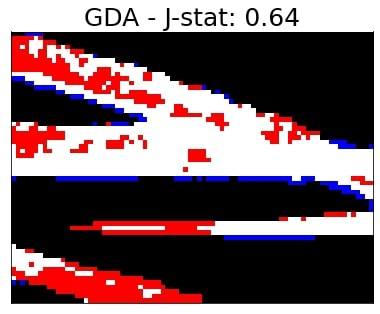}
        \includegraphics[scale = 0.35]{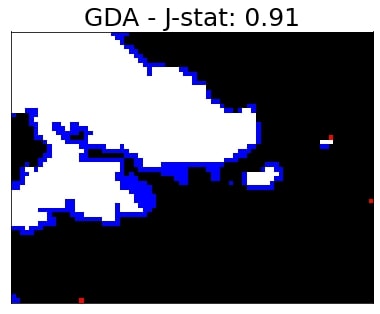}
        \includegraphics[scale = 0.35]{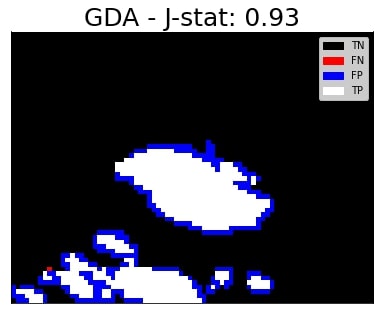}
    \end{minipage}
    \begin{minipage}{\linewidth}
        \centering
        \includegraphics[scale = 0.35]{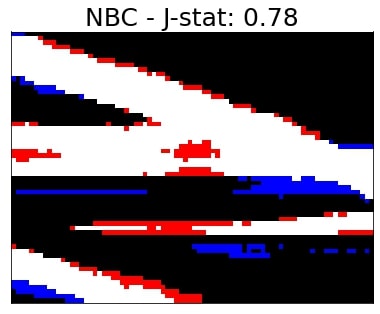}
        \includegraphics[scale = 0.35]{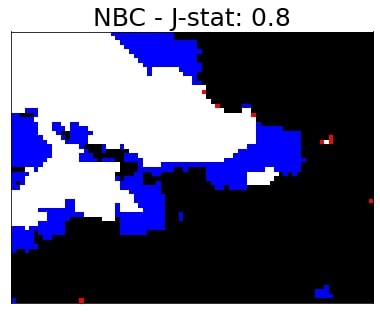}
        \includegraphics[scale = 0.35]{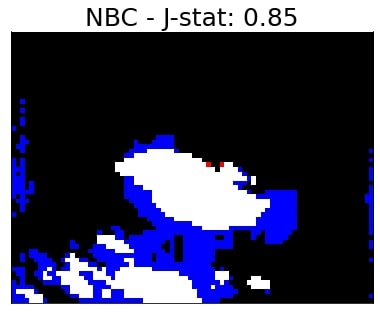}
        \end{minipage}
    \begin{minipage}{\linewidth}
        \centering
        \includegraphics[scale = 0.35]{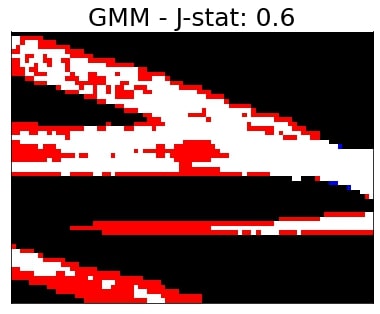}
        \includegraphics[scale = 0.35]{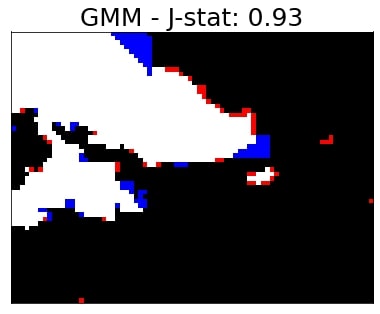}
        \includegraphics[scale = 0.35]{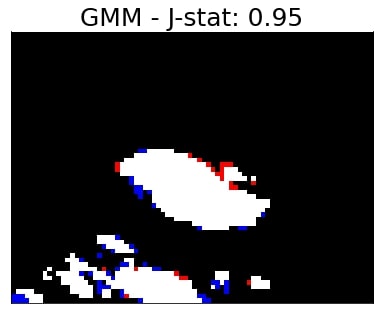}
    \end{minipage}
    \begin{minipage}{\linewidth}
        \centering
        \includegraphics[scale = 0.35]{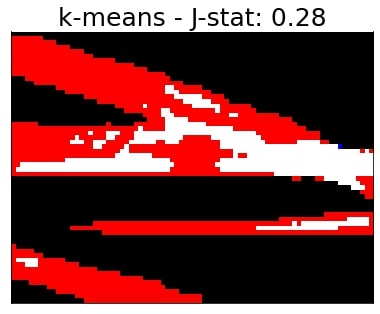}
        \includegraphics[scale = 0.35]{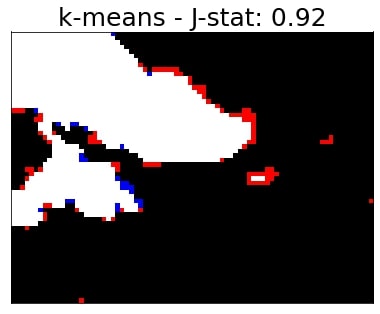}
        \includegraphics[scale = 0.35]{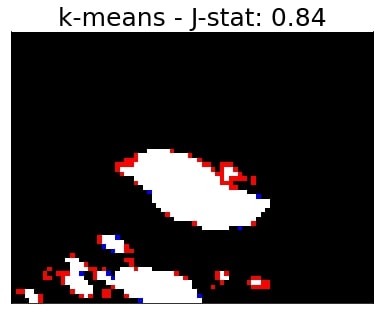}
    \end{minipage}
\caption{Three images from the test organized in columns. The images in each row show the segmentation performed by a generative model. The higher j-statistic was achieved by the NBC in the first image, and the GMM in the second and third images.}
\label{fig:generative_test}
\end{figure}

\begin{figure}[!htb]
    \begin{subfigure}{\linewidth}
        \centering
        \includegraphics[scale = 0.35]{images_v1/labels_1.jpg}
        \includegraphics[scale = 0.35]{images_v1/labels_2.jpg}
        \includegraphics[scale = 0.35]{images_v1/labels_3.jpg}
    \end{subfigure}
    \begin{minipage}{\linewidth}
        \centering
        \includegraphics[scale = 0.35]{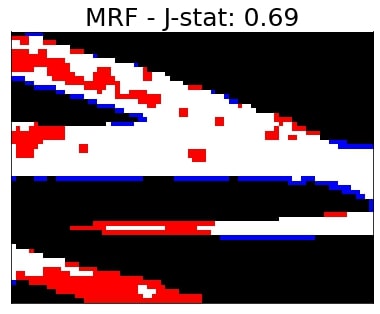}
        \includegraphics[scale = 0.35]{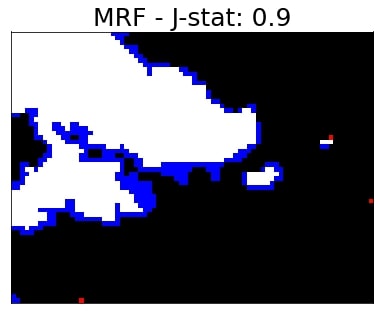}
        \includegraphics[scale = 0.35]{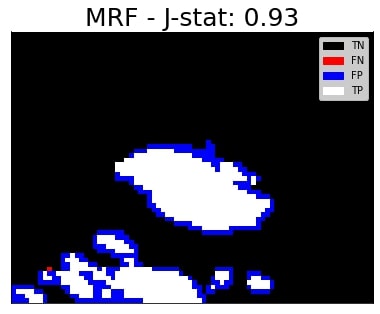}
    \end{minipage}
    \begin{minipage}{\linewidth}
        \centering
        \includegraphics[scale = 0.35]{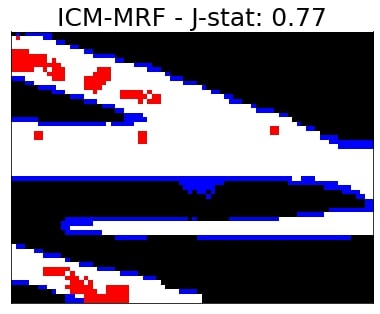}
        \includegraphics[scale = 0.35]{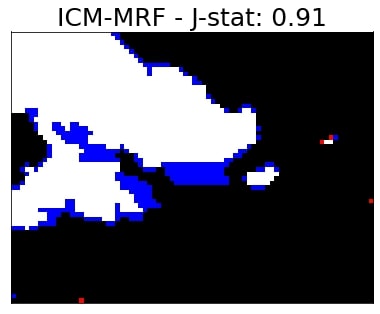}
        \includegraphics[scale = 0.35]{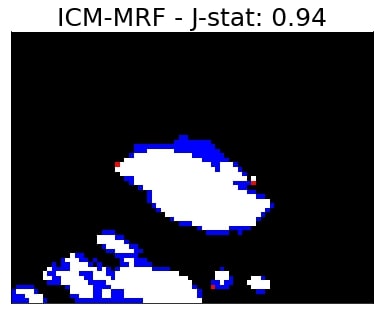}
    \end{minipage}
    \begin{minipage}{\linewidth}
        \centering
        \includegraphics[scale = 0.35]{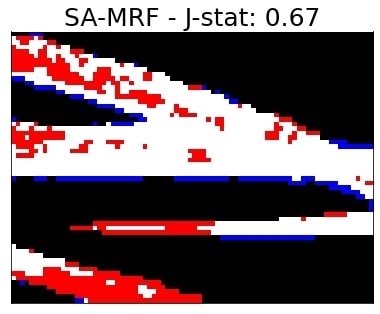}
        \includegraphics[scale = 0.35]{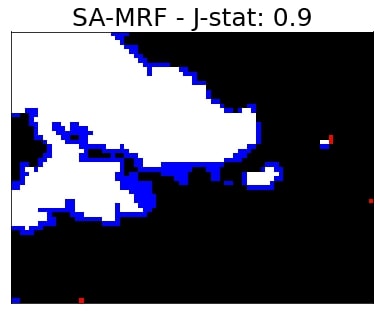}
        \includegraphics[scale = 0.35]{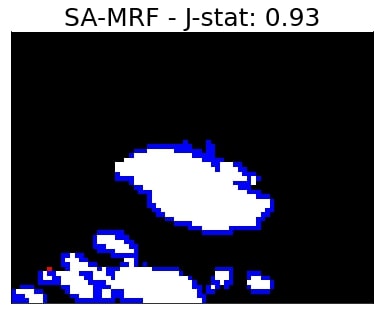}
    \end{minipage}
    \begin{minipage}{\linewidth}
        \centering
        \includegraphics[scale = 0.35]{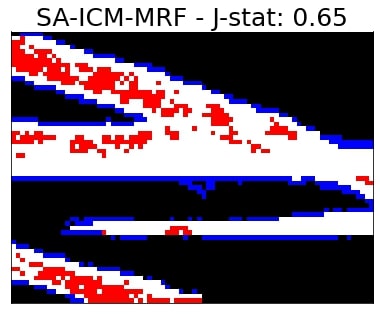}
        \includegraphics[scale = 0.35]{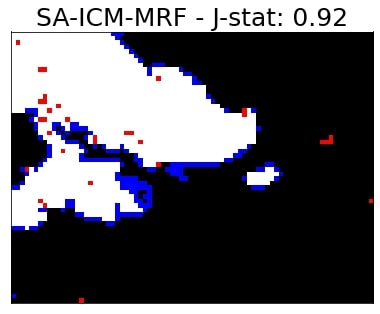}
        \includegraphics[scale = 0.35]{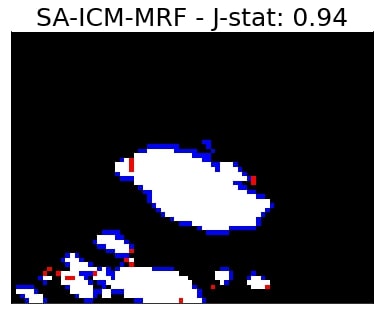}
    \end{minipage}
\caption{Three images from the test subset organized in columns. The images in each row show the segmentation performed by a MRF model. The highest j-statistic was achieved by ICM-MRF in the first image, SA-ICM-MRF in the second, and ICM-MRF and SA-ICM-MRF in the third image.}
\label{fig:mrf_test}
\end{figure}

\begin{figure}[!htb]
    \begin{subfigure}{\linewidth}
        \centering
        \includegraphics[scale = 0.35]{images_v1/labels_1.jpg}
        \includegraphics[scale = 0.35]{images_v1/labels_2.jpg}
        \includegraphics[scale = 0.35]{images_v1/labels_3.jpg}
    \end{subfigure}
    \begin{minipage}{\linewidth}
        \centering
        \includegraphics[scale = 0.35]{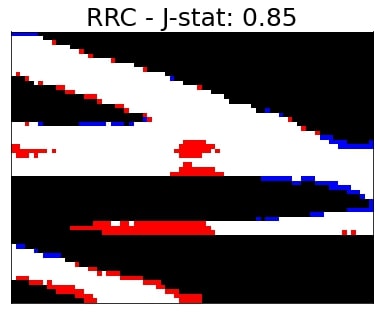}
        \includegraphics[scale = 0.35]{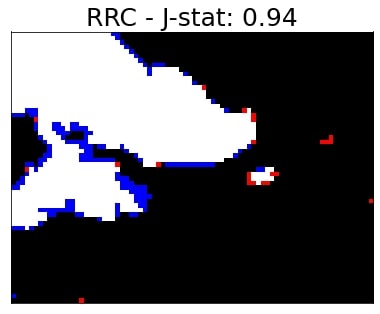}
        \includegraphics[scale = 0.35]{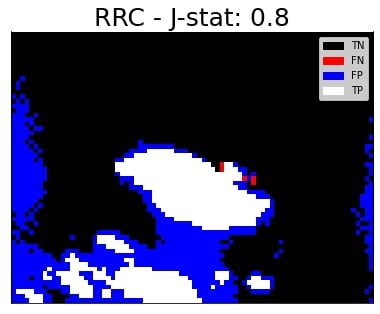}
    \end{minipage}
    \begin{minipage}{\linewidth}
        \centering
        \includegraphics[scale = 0.35]{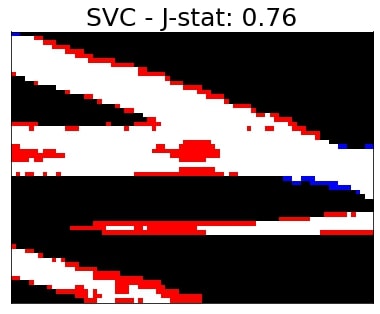}
        \includegraphics[scale = 0.35]{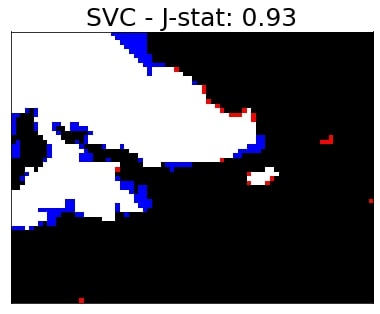}
        \includegraphics[scale = 0.35]{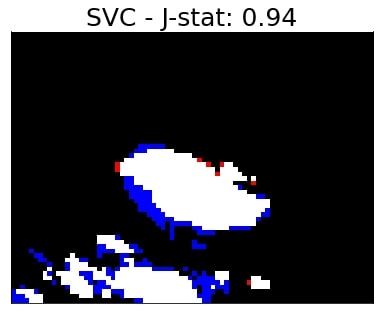}
    \end{minipage}
    \begin{minipage}{\linewidth}
        \centering
        \includegraphics[scale = 0.35]{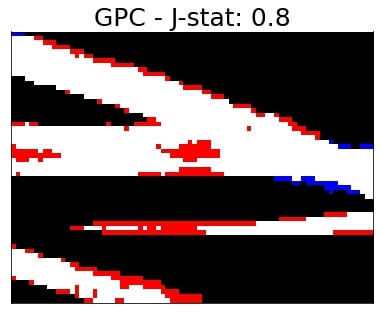}
        \includegraphics[scale = 0.35]{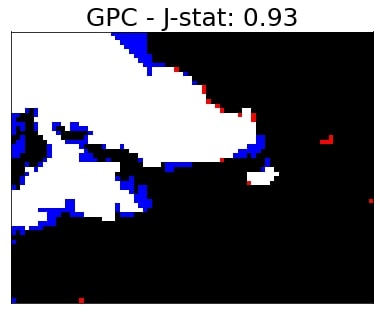}
        \includegraphics[scale = 0.35]{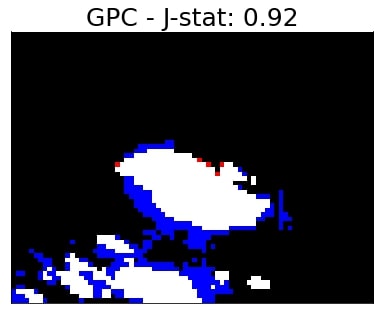}
    \end{minipage}
\caption{Three images from the test subset are organized in columns. The images in each row show the segmentation performed by the discriminative models. When segmenting the images, a higher j-statistic was achieved by RRC (in the first and second image), and SVC (in the third image).}
\label{fig:discriminative_test}
\end{figure}

The segmentation performed on three testing images by the generative models are shown in Fig. \ref{fig:timing}-\ref{fig:mrf_test}. The NBC and GDA are both discriminant analysis and supervised learning models (Fig. \ref{fig:generative_test}). The k-means and GMM are unsupervised learning methods (Fig. \ref{fig:generative_test}). The MRF and SA-MRF are supervised learning models and ICM-MRF and SA-ICM-MRF are unsupervised learning models. The SA algorithm is implemented to speed-up the MRF and ICM-MRF convergence. When MRF models use the SA algorithm, the segmentation is not so uniform (Fig. \ref{fig:mrf_test}). The cooling mechanism of the SA algorithm ends the optimization before the segmentation has converged to a state of higher energy. The discriminative models used are the RRC, SVC and GPC (Fig. \ref{fig:discriminative_test}). These were solved in the primal formulation so their performances are feasible for real-time cloud segmentation (see Fig. \ref{fig:timing}). The performances of the models are compared in terms of j-statistic vs. training computing time vs. average computing time in testing. The j-statistic is evaluated with the images in the testing subset. The computing time is measured in seconds. The time in the y-axes of the graphs shown in Fig. \ref{fig:timing} are displayed in logarithmic scale. The highest j-statistic is achieved by the unsupervised MRF, but the training and the average testing computing time are the largest. NBC and RRC have the lowest training times. In the implementation, the k-means, NBC and RRC have the lowest computing time. If we have considered all of this information, the most suitable model would be one of these model.

The unsupervised MRF model (ICM-MRF) achieved the highest j-statistic in testing among generative and discriminative models. The ICM-MRF model uses the feature vector $\mathbf{x}^{3}$ with a $1^{st}$ order neighborhood and the set of cliques $\Omega_{1}$ in the prior. The classification performance of the model decreased when optimized using the SA algorithm, but the average testing time was faster (Fig. \ref{fig:mrf}). The MRF models that use a prior potential function lead to the largest training and average testing computational time. When only generative models without the prior potential function are considered (NBC, GDA, k-means and GMM), the GDA has the highest j-statistic with the feature vector $\mathbf{x}^{4}$ of a $2^{nd}$ order neighborhood (Fig. \ref{fig:generative}). However, if the trade-off between average testing time and j-statistic is considered, the most suitable generative model is the GMM with a feature vector $\mathbf{x}^{4}$ of a single pixel neighborhood. The generative models which include a simplification of the covariance matrix and that do not use a prior potential function (NBC and k-means) yield the fastest average testing time among all classification models (without considerably decreasing the j-statistic). However, these models have the lowest j-statistic among all the models implemented. Fig. \ref{fig:discriminative} shows the discriminative models' j-statistics. The average testing time is lower than that obtained by the generative MRF models, but the j-statistic is higher than that obtained by the generative models without the potential function. The polynomial expansion yields overfitting in all the discriminative models. The discriminative model that achieved the highest j-statistic is the linear GPC with the feature vector $\mathbf{x}^{4}$ of a single pixel neighborhood. As seen in Fig. \ref{fig:timing}, the RRC and SVC are the most suitable methods, as they offer the best compromise between average testing times and accurate segmentation.

\begin{figure}[!htb]
   \begin{subfigure}{0.325\linewidth}
        \includegraphics[scale = 0.35]{images_v1/labels_1.jpg}
    \end{subfigure}
    \begin{subfigure}{0.325\linewidth}
        \includegraphics[scale = 0.35]{images_v1/labels_2.jpg}
    \end{subfigure}
    \begin{subfigure}{0.325\linewidth}
        \includegraphics[scale = 0.35]{images_v1/labels_3.jpg}
    \end{subfigure}
    \centering
    \begin{subfigure}{0.325\linewidth}
        \includegraphics[scale = 0.35]{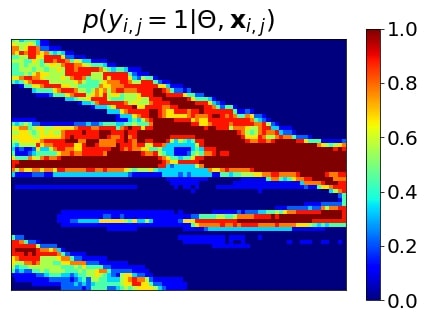}
    \end{subfigure}
    \begin{subfigure}{0.325\linewidth}
        \includegraphics[scale = 0.35]{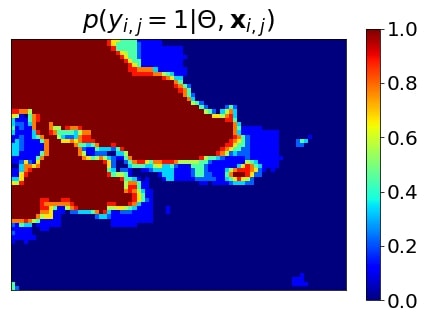}
    \end{subfigure}
    \begin{subfigure}{0.325\linewidth}
        \includegraphics[scale = 0.35]{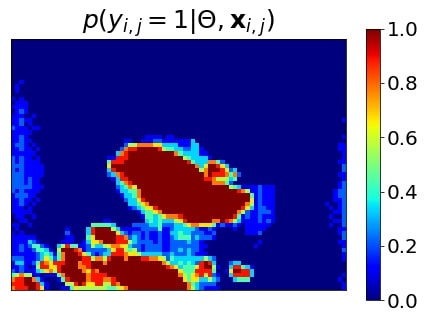}
    \end{subfigure}
   \begin{subfigure}{0.325\linewidth}
        \includegraphics[scale = 0.35]{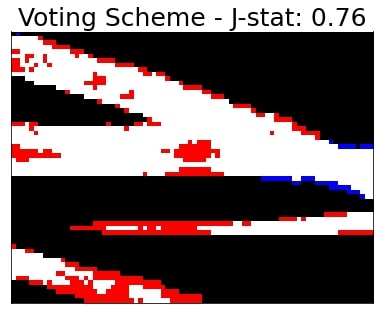}
    \end{subfigure}
    \begin{subfigure}{0.325\linewidth}
        \includegraphics[scale = 0.35]{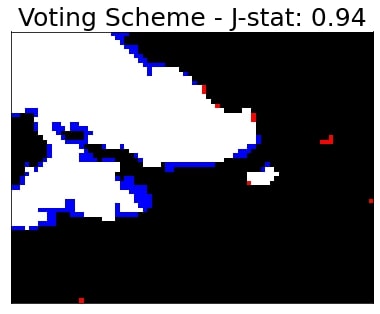}
    \end{subfigure}
    \begin{subfigure}{0.325\linewidth}
        \includegraphics[scale = 0.35]{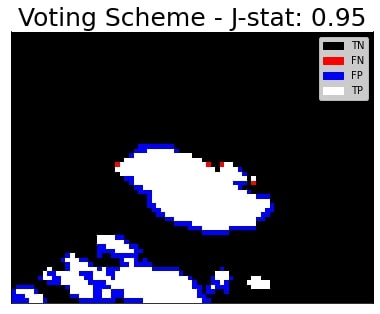}
    \end{subfigure}
   \begin{subfigure}{0.325\linewidth}
        \includegraphics[scale = 0.35]{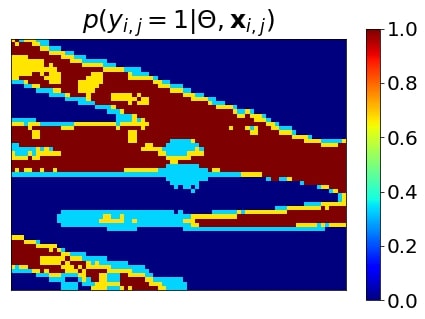}
    \end{subfigure}
    \begin{subfigure}{0.325\linewidth}
        \includegraphics[scale = 0.35]{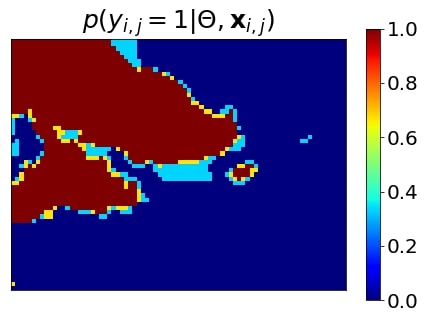}
    \end{subfigure}
    \begin{subfigure}{0.325\linewidth}
        \includegraphics[scale = 0.35]{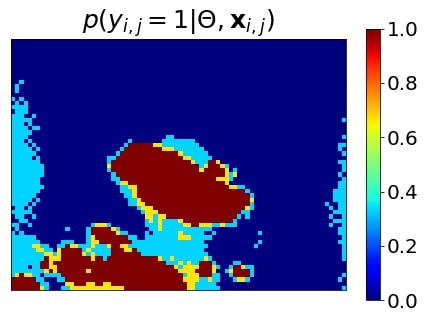}
    \end{subfigure}
   \begin{subfigure}{0.325\linewidth}
        \includegraphics[scale = 0.35]{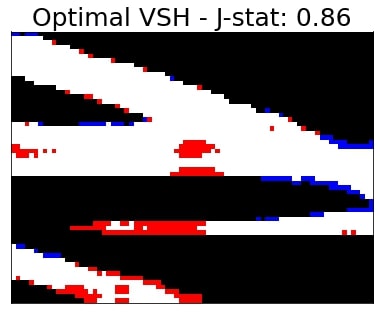}
    \end{subfigure}
    \begin{subfigure}{0.325\linewidth}
        \includegraphics[scale = 0.35]{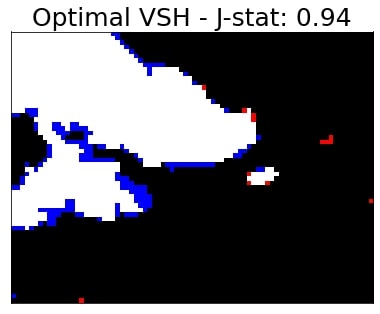}
    \end{subfigure}
    \begin{subfigure}{0.325\linewidth}
        \includegraphics[scale = 0.35]{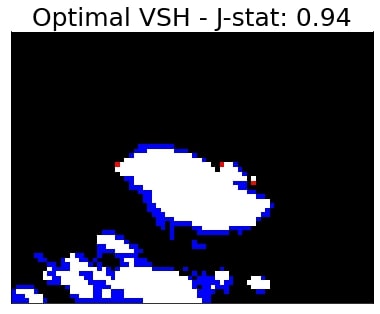}
    \end{subfigure}
\caption{Three different test images. First and second rows:  results of the voting scheme. The first row displays the probability of a pixel belonging to a cloud. The second row shows the segmentation performed by the voting scheme. Third and fourth rows:  probability of a pixel belonging to a cloud and the segmentation of the optimal voting scheme (VSH).}
\label{fig:voting_test}
\end{figure}

The results in Fig. \ref{fig:generative}-\ref{fig:discriminative} show the importance of the feature extraction method in cloud image segmentation. The extraction of features makes it easier for the models to differentiate between cloudy and clear-sky pixels, because the distance between feature vectors of different classes increases in the feature space. Through feature extraction, the feature vectors of the same classes group together forming clusters in the feature space. Without extracting features correctly, the feature vectors from both classes (cloudy and clear-sky) are grouped in a single cluster, making it difficult to perform a classification. When the magnitude of velocity vectors are included in the feature vectors, combined with temperature increments and normalized temperature increments, the segmentation models achieved a higher j-statistic. The addition of features from neighboring pixels to the feature vectors improves the performance in some of the models. 

When the raw temperature and height are used, all models have poor performance. However, when the images are preprocessed with the outdoor germanium camera window model and the atmospheric model, the ICM-MRF reaches a reasonable performance of 92.55 \% at the expense of a high computational cost of 641 ms per image in testing. The performance of discriminative methods with this set of features is lower, ranging between 72.58 \% and 84 \%. When velocity vectors are added to the features, the discriminative methods achieve a similar performance as the ICM-MRF, with computational times of 2.2 ms (RRC), 3.7 ms (SVC) and 77 ms (GPC). The best compromise is the SVC, which is 150 times faster than the ICM-MRF with a small difference in accuracy. The image preprocessing and feature extraction time is 0.1 ms for ${\bf x}^1$, 4.7 ms for ${\bf x}^2$, 99.9 ms for ${\bf x}^3$ and 1079 ms for ${\bf x}^4$. When preprocessing time is added to the segmentation time, the average time required by the ICM-MRF is 740.9 ms. This is faster than the average time required by the discriminative models: 1081 ms (RRC), 1083 ms (SVC) and 1156 ms (GPC).

A voting scheme using the predictions from the models displayed in Fig. \ref{fig:discriminative_test}-\ref{fig:discriminative_test} (not including SA-MRF and SA-ICM-MRF) achieved higher j-statistic but have a higher computing time. The j-statistic is 93 \%, see Fig. \ref{fig:voting_test}. The combination of the RRC, SVC and ICM-MRF lead to the best j-statistic . The optimal voting scheme reached a j-statistic of 94.68 \% in testing (see Fig. \ref{fig:voting_test}). The voting scheme's training and testing times are the sum of each method's respective computing times. When the models are trained and tested in parallel, the voting scheme's training and testing times are that of the slower models.

\section{Conclusion}

This investigation seeks to find the optimal methods for real-time ground-based IR cloud segmentation through image preprocessing and feature extraction. Preprocessing was applied to remove underlying cycle-stationary processes, and feature extraction was used to compute cloud height and velocity. The results show that cloud segmentation in ground-based IR images is not only feasible, but achieves high performance in real-time applications. Ground-based IR cameras perform better than visible ones in poor light conditions. We implement a prepocessing algorithm that uses physical features extracted from IR images. The j-statistic is proposed to independently measure the accuracy of the classification in each classes.

Preprocessing the ground-based IR images using the window and atmospheric models leads to an overall performance improvement. Simplification of the covariance matrix reduces the computing time, but the j-statistic achieved is lower than that of the models using the full covariance matrix. Adding the features of neighboring pixels to the feature vectors yields an increase in segmentation performance in some cases. The discriminative models formulated in the primal result in feasible segmentation models for real-time application. MRF models remove possible outliers using cliques from neighboring pixels. This increases the overall performance of the generative models when trained with unsupervised and supervised algorithms. The optimal voting scheme achieved the best j-statistic. However, the implementation computing time might be slow for real-time applications when not run in parallel.

Further investigations may focus on segmentation in multiple layers of clouds. The clouds in each layer may be segmented into different classes. An algorithm can be trained to detect multiple layers of clouds when clouds have different heights or  directions. In this way, the extraction of features may be performed independently in each one of the cloud layers. A multiple cloud layer segmentation algorithm will reduce the noise when extracting features. This algorithm may be implemented to increase the performance of ground-based intra-hour GSI forecasting.

\section{Acknowledgments}

Partially supported by NSF EPSCoR grant number OIA-1757207 and the King Felipe VI endowed Chair. Authors  thank the UNM Center for Advanced Research Computing, supported by the NSF, for providing the high performance computing and large-scale storage resources.
\bibliographystyle{unsrt}  
\bibliography{mybibfile}

\end{document}